\documentclass[final,leqno,onefignum,onetabnum]{siamltex704}
\usepackage{url}
\usepackage{graphics,graphicx,epsf,subfigure,epstopdf}
\usepackage{times}
\usepackage{multirow}
\usepackage{amsmath,amsxtra,amsfonts,amscd,amssymb,bm}
\usepackage[algo2e,linesnumbered,vlined,ruled]{algorithm2e}
\usepackage{caption}
\usepackage{algorithm,lipsum,changepage}
\usepackage[noend]{algpseudocode}
\newcommand{\be}{\begin{equation}}
\newcommand{\ee}{\end{equation}}
\newcommand{\bee}{\begin{equation*}}
\newcommand{\eee}{\end{equation*}}
\newcommand{\bea}{\begin{eqnarray}}
\newcommand{\eea}{\end{eqnarray}}
\newcommand{\beaa}{\begin{eqnarray*}}
\newcommand{\eeaa}{\end{eqnarray*}}

\newcommand{\st}{\,\mathbf{s.t.}\,}
\newcommand{\Z}{\mathbb{Z}}

\newcommand{\B}{\mathcal{B}}

\begin{document}
\title{Randomized Algorithms For High Quality Treatment Planning in Volumetric Modulated Arc Therapy}
%\author{Yu Yang, Zaiwen Wen, Bin Dong}
%\thanks{Yu Yang (1200010726@pku.edu.cn) is with the School of Mathematical Sciences, Peking University; Zaiwen Wen (wenzw@pku.edu.cn) and Bin Dong (dongbin@math.pku.edu.cn) are with Beijing International Center for Mathematical Research (BICMR), Peking University.}
%\thanks{Corresponding author(s): Zaiwen Wen and Bin Dong}
%\thanks{\textcolor{blue}{Yu Yang and Zaiwen Wen are supported by xxxx}; Bin Dong is supported by the Thousand Talents Plan of China.}
%

\author{Yu Yang\footnotemark[1] \and
Bin Dong\footnotemark[2] \and
Zaiwen Wen\footnotemark[3]}

\renewcommand{\thefootnote}{\fnsymbol{footnote}}
\footnotetext[1]{School of Mathematical Sciences, Peking University, Beijing, CHINA (1200010726@pku.edu.cn).}
\footnotetext[2]{Corresponding author. Beijing International Center for Mathematical
Research, Peking University, Beijing, CHINA (dongbin@math.pku.edu.cn).
Research supported in part by the Thousand Talents Plan of China.}

\footnotetext[3]{Beijing International Center for Mathematical
Research, Peking University, Beijing, CHINA (wenzw@pku.edu.cn).
Research supported in part by NSFC grant 11322109 and by the National
Basic Research Project under the grant 2015CB856000.}

\maketitle
\begin{abstract}
In recent years, volumetric modulated arc therapy (VMAT) has been becoming a more and more important
radiation technique widely used in clinical application for cancer treatment. One of the key problems in VMAT is treatment plan optimization,
which is complicated due to the constraints imposed by the equipments involved. In this paper, we consider a model with four major constraints: the
bound on the beam intensity, an upper bound on the rate of the change of the
beam intensity, limit on the moving speed of leaves of the multi-leaf collimator (MLC) and its directional-convexity. We solve the model by a two-stage
algorithm: performing minimization with respect to the shapes of the aperture and the beam intensities alternatively. Specifically, the shapes of the aperture are obtained by a greedy algorithm whose performance is enhanced by random sampling in the leaf pairs with a decremental rate. The beam intensity is optimized using a gradient projection method with nonmonotonic line search. We further improve the proposed algorithm by an incremental random importance sampling of the voxels to reduce the computational cost of the evaluation of the energy function.  Numerical simulations on two clinical data sets demonstrate that our method is highly competitive to the state-of-the-art algorithms in terms of both computation time and quality of treatment planning.
\end{abstract}
\begin{keywords}  Volumetric Modulated Arc Therapy, Greedy algorithm, Gradient projection, Random sampling, Importance sampling.
\end{keywords}

\section{Introduction}

Cancer is one of the most deadly diseases, causing millions of deaths all over the world every year. According to the World Cancer Report by the World Health Organization in 2014, about 14.1 million new cases of cancer occurred globally in 2012. It caused about 8.2 million deaths or 14.6\% of all human deaths. The data from United States National Cancer Institute indicates that an estimated 1,658,370 new cases of cancers is diagnosed in the United States and 589,430 people die from these diseases in 2015. Therefore, cancer prevention, diagnosis and treatment are of hyper importance to the world.

\begin{figure}[htb]
\setcaptionwidth{\linewidth}
\centering
\subfigure{
\begin{minipage}[t]{0.4\linewidth}
\centering
\includegraphics[width=0.8\textwidth]{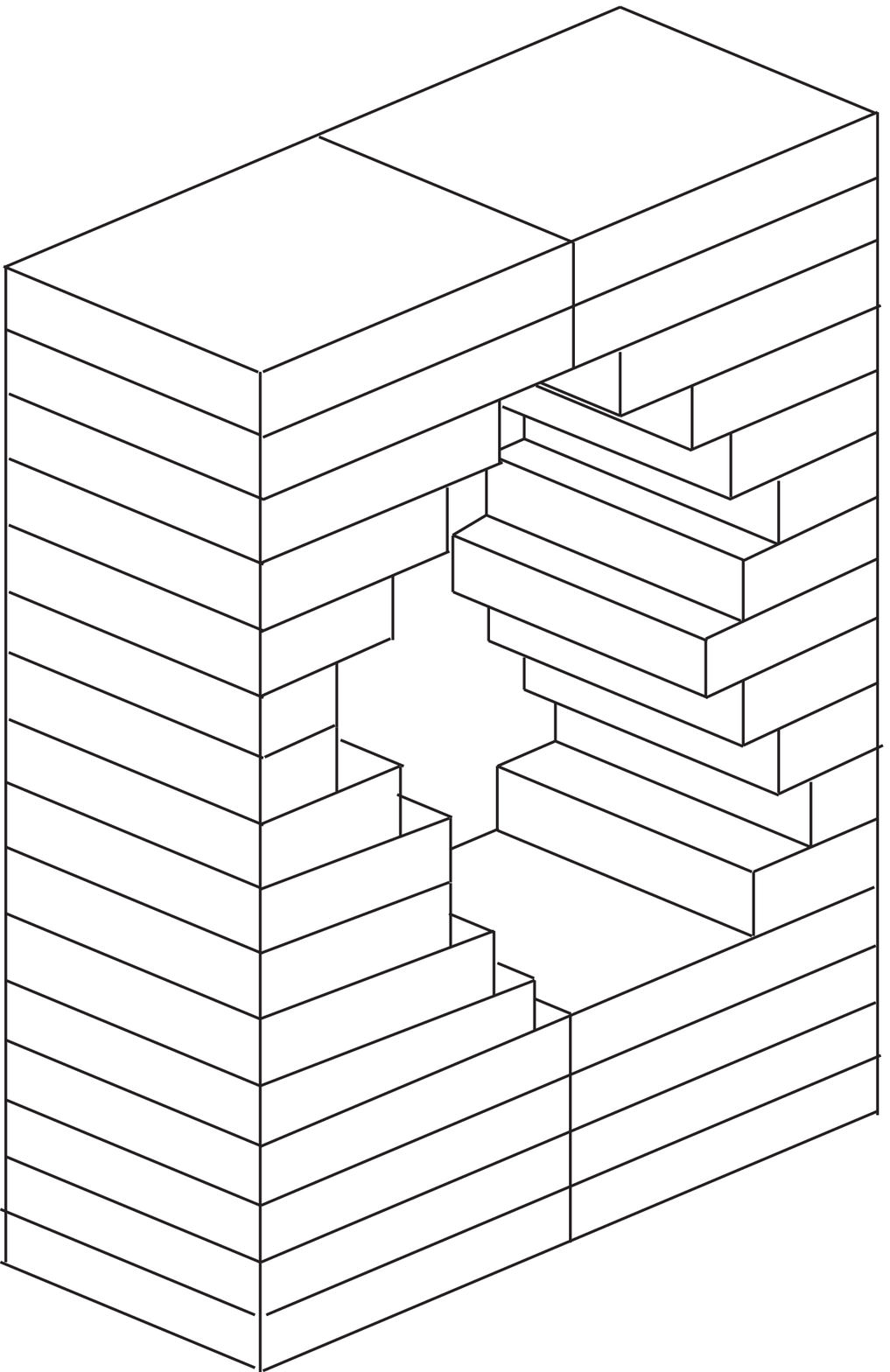}
\end{minipage}
\begin{minipage}[t]{0.4\linewidth}
\centering
\includegraphics[width=\textwidth]{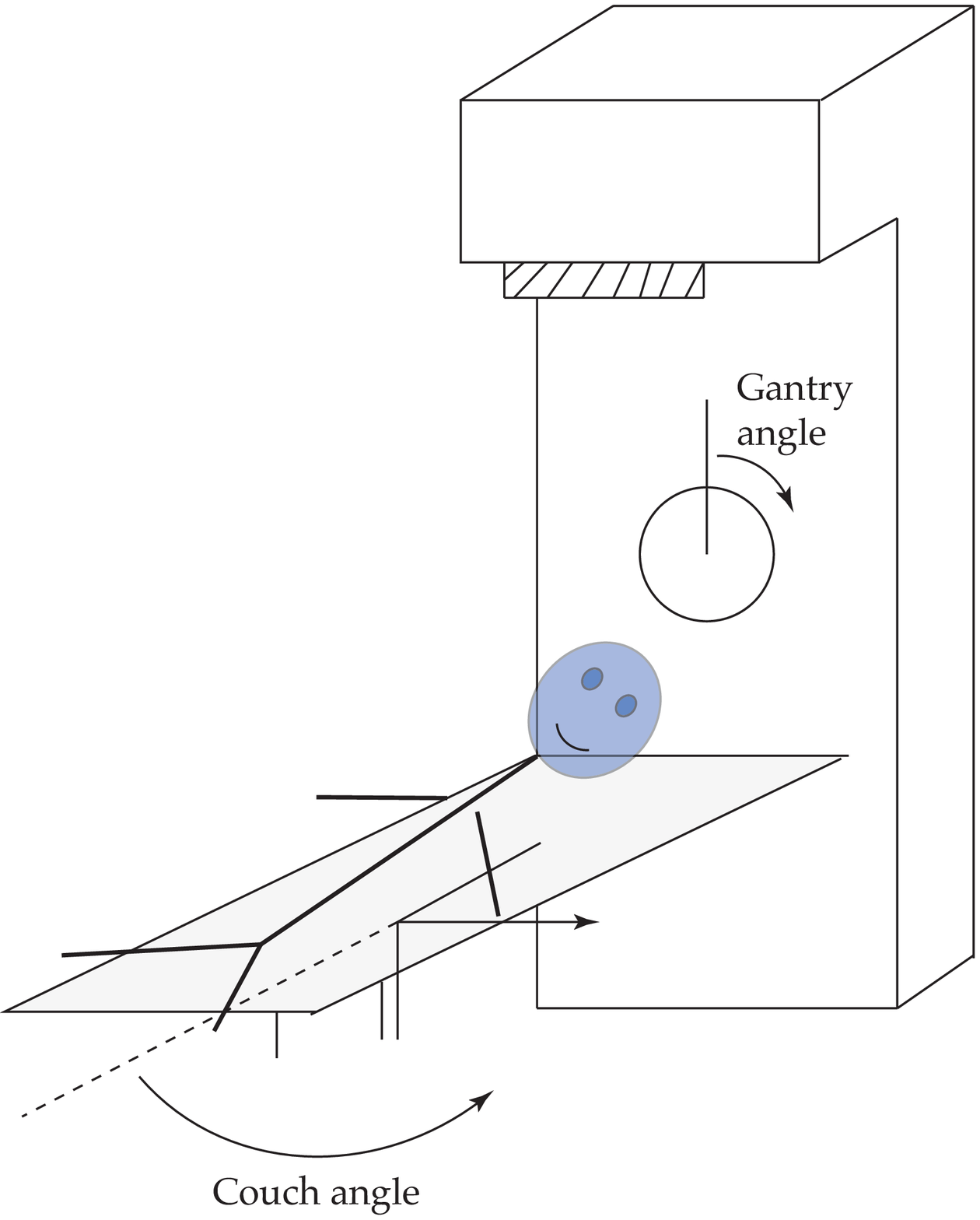}
\end{minipage}}
\caption{Left: Diagram of a MLC  \cite{Dong2012}. Right: The definition of the gantry and the
couch angle \cite{Data2014}.}\label{figure 1}
\end{figure}

Radiation therapy is frequently used in cancer treatment. It is commonly applied to the cancerous cells because of its ability to control cell growth. Most common cancer types can be treated with radiation therapy to a certain extent. It uses relatively high-energy doses of ionizing radiation to damaging the DNA of cancerous tissues leading to cellular death. In the process of the therapy, a radiation beam is generated from a medical linear
accelerator fixed in a gantry that can rotate around the patient so that tumors in the patient can receive radiation from various directions. At each direction, the beam is collimated into desired shapes through a device called multi-leaf collimator (MLC) before hitting the patient. An example of
MLC is shown in the left side of Figure \ref{figure 1}. The leaves of MLC can only move in a certain direction at a certain speed. Furthermore, the beam intensity can only vary within a certain range and change at a certain rate. These mechanical limitations impose a few difficult constraints
in designing a suitable treatment plan.

Intensity Modulated Radiation Therapy (IMRT) and Volumetric-Modulated Arc
Therapy (VMAT) are two major techniques for radiation therapy. IMRT was proposed by Cedric Yu  in
\cite{Yu1995} as an alternative to tomotherapy and much progress \cite{Wang,Brahme,Burman} has been made over the years. In IMRT, a few angles are selected in advance and multiple shapes of the aperture at each angle are used. In VMAT, on the other hand, the beam intensities and shapes of the aperture change continuously and radiation can be delivered to the targets during the whole rotation of the gantry \cite{otto2008volumetric}. For some special cases, the couch on which the patient lies can also rotate to fully take advantage of the potentials of VMAT (see Figure \ref{figure 1}). Therefore, VMAT is able to significantly reduce the treatment time compared to IMRT without sacrificing treatment quality \cite{Palma,Verbaket}. Hence, it has experienced an increasing popularity in clinical application \cite{Tang} in the last few years. However, since VMAT is much more complicated than IMRT because of the flexibility and additional constraints due to mechanical limitations, further improvements of current treatment planning algorithms for VMAT are still needed.

Several models and algorithms have already been developed for the treatment planning in VMAT. The key issue, which is also the main challenge, is to achieve a good balance between delivering sufficient dose to tumors and causing minimal damage to normal tissues and organs. A proper energy functional is often designed based on certain prior information on the patient's anatomy and it offers a proper compromise between the desired dose at the target and the dose at healthy organs. Furthermore, one also needs to consider constraints on the bounds of the beam intensity, an upper bound on the rate of the change of the
beam intensity, limit on the moving speed of leaves of the multi-leaf collimator (MLC) and its directional-convexity. An integer programming problem was formulated in \cite{akartunalitreatment} and four methods are proposed: two heuristic strategies based on the Lagrangian relaxation, one heuristic scheme based on a reformulation of the problem, and a metaheuristic strategy based on the guided variable neighborhood scheme. All of these methods can be computationally expensive. A new column-generation-based algorithm was proposed in \cite{Peng} which was implemented using GPU which significantly improved the efficiency. Nevertheless, their model over-simplified the problem and only a limited number of apertures were considered. In \cite{BofeiSun2013}, the authors formulated the problem as a nonlinear integer programming problem , and a two-stage algorithm is proposed. More recently, an alternating minimization framework was developed in \cite{Dong2012}. The authors use a level-set strategy to represent beam shapes, and a fast sweeping technique is applied to calculate dose intensity. Although their model incorporates most of the main features of VMAT, the computation efficiency of their algorithms still needs to be improved.

In this paper, we propose a mixed-integer nonlinear and nonconvex model. The structures in the patient's body are classified into three categories and they are treated differently according to their relative importance. The objective function is constructed as a combination of the quadratic and cubic function, which enables us to achieve a good balance between sufficient dose delivery to the targets and protection of healthy tissues and organs. Due to the physical
constraints on the movement of the leaves of MLC, the aperture shapes are characterized by integer variables to describe the directional convexity constraint.
Since it is difficult to determine the shapes of the aperture and beam intensities simultaneously, we adopt an iterative alternating minimization framework which solves the model with respect to one variable while the other variable is fixed. The subproblem that determines the shapes of the aperture is combinatorial and is computed using a greedy strategy. Essentially, it revises the boundary of the shapes locally and the elements to be updated are fixed at each iteration. The subproblem that calculates the beam intensities is differentiable and is solved by a standard gradient-projection method using nonmonotone line search with the BB step size. The most computationally intense part of our algorithms is calculating the dose distribution and evaluating the energy function. They are computationally expensive because the number of voxels in the discretization of the energy functional can be up to several millions.  Hence, we propose an incremental randomized sampling strategy which only takes a small proportion of the voxels in the energy function and the selection is based on the importance of the voxels. A decremental scheme is also developed to update the boundary of the aperture shapes in the greedy algorithm. Numerical experiments on the prostate and head-and-neck cancer cases in a real medical dataset show that our algorithms can solve the problem more accurately and efficiently than the state-of-the art algorithms. In some cases, the randomized strategy can significantly outperform the deterministic scheme.

The remainder of this paper is organized as follows. In Section \ref{sec:model}, we introduce a mathematical model for VMAT and present its discretized form. In Section \ref{sec:alg}, we develop a greedy algorithm for finding the shapes of the aperture and propose a gradient projection method to optimize beam intensities. The two steps are performed alternately until convergence. Incremental random samplings are introduced to further enhance the performance. Our numerical experiments are demonstrated in Section \ref{sec:num}. Finally, we conclude the paper in Section \ref{sec:conclusion}.

\section{The VMAT Optimization Model} \label{sec:model}

\subsection{The Energy Functional}
\begin{figure}[H]
\centering
\subfigure{
\begin{minipage}[t]{0.4\linewidth}
\centering
\includegraphics[width=\textwidth]{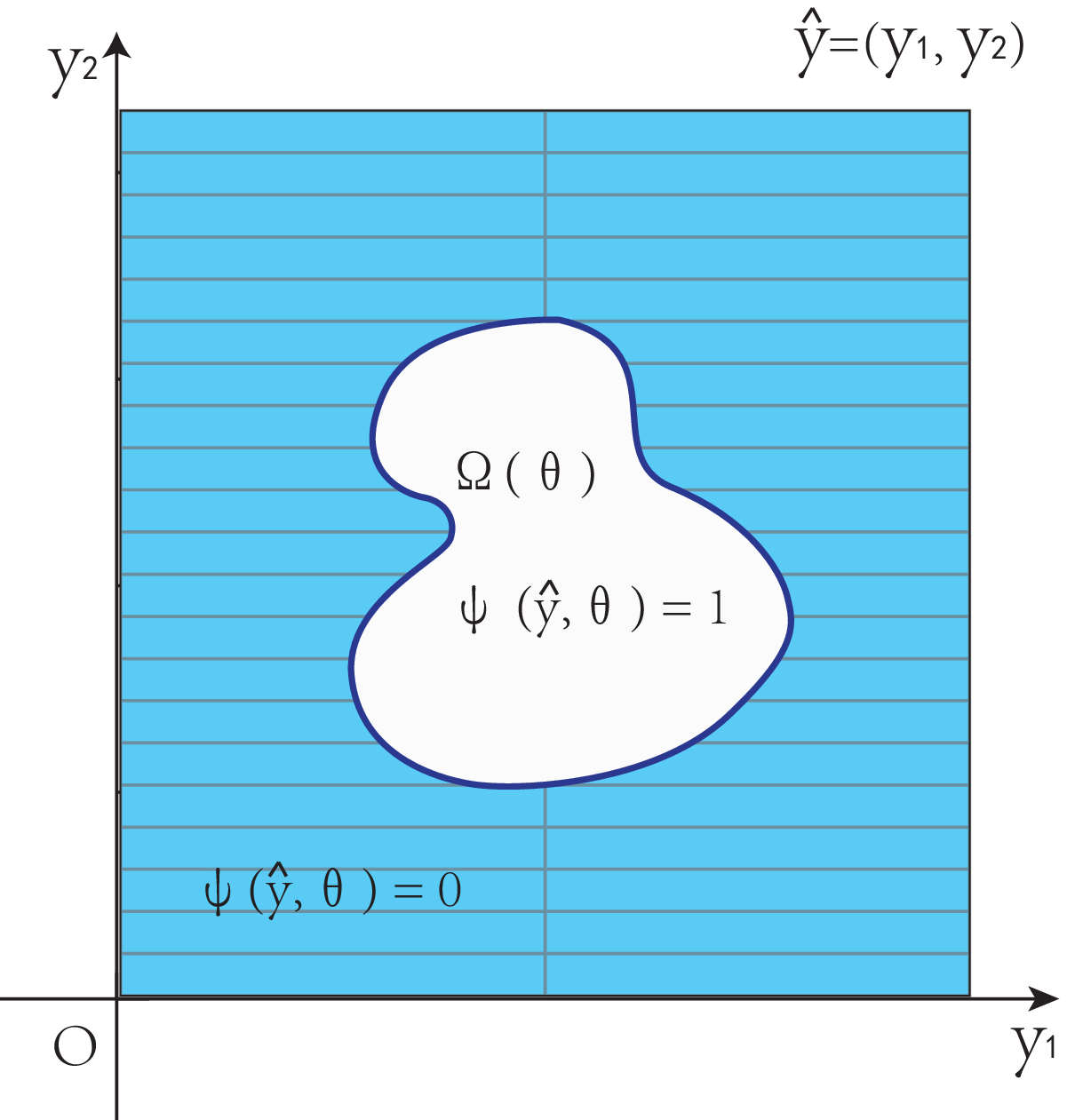}
\end{minipage}
\begin{minipage}[t]{0.4\linewidth}
\centering
\includegraphics[width=0.92\textwidth]{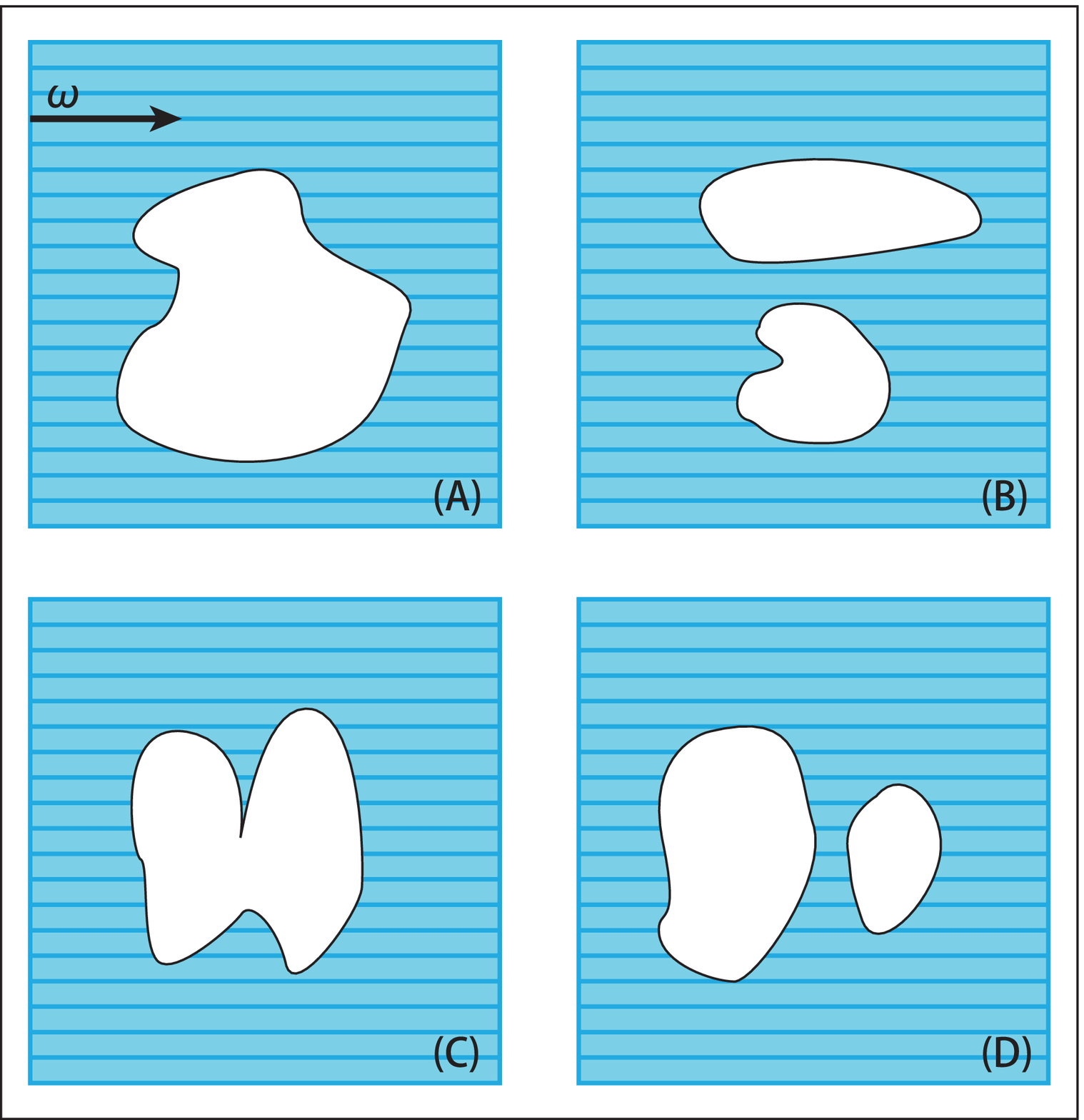}
\end{minipage}}
\caption{Left: An aperture at a given gantry angle $\theta$ described by the indicator function $\psi $.
Right: An illustration of deliverable apertures ((A) and (B)) and undeliverable ones ((C) and (D))
when leaves are oriented in the ${\omega}$ direction}\label{fig:2}
\end{figure}

%We first introduce an objective function in order to solve VMAT from a perspective of optimization.
In VMAT, radiation is delivered continuously to the patient in a single rotation. Let the gantry angle $\theta \in[0,2\pi ]$ denote the position of the gantry. For some cases, the couch where the patient lies can also rotate during the procedure. The couch angle can only be chosen in the range of $[-\frac\pi2,\frac\pi2]$. Each time the couch moves to a selected angle, the gantry begins to rotate. Consequently, the entire process can be simplified by assuming that the couch is fixed while the gantry rotates more than one circle. In this case, the angle $\theta$ does not necessarily fall into the range of $[0,2\pi]$. Let $s(\theta)$ denote the beam intensity at angle $\theta$, $\Omega(\theta)$ be the aperture formed by the MLC at that angle, and $\hat y\in\mathbb{R}^2$ be a location in the MLC plane. Then, the dose distribution $z(x)$, with $x\in\mathbb{R}^3$, takes the form:
\[z(x)=\int_0^{2\pi} \int_{\Omega(\theta)} D(x,\hat y,\theta)s(\theta) \mathrm{d}\hat y \mathrm{d}\theta,\]
where $D(x,\hat y,\theta)$ is the dose-influence coefficient indicating the dose received at location $x$ in the patient's body when per unit intensity of beamlet radiation is delivered through location $y$ in the MLC plane where the aperture is open. The coefficient $D(x,\hat y,\theta)$ is generated from CERR (Computational Environment for Radiotherapy Research) beforehand, and is computed specifically for each patient. The aperture shape $\Omega(\theta)$ created by the MLC can be represented equivalently by an indicator function $\psi$ whose value is equal to $1$ inside the shape $\Omega(\theta)$ and
$0$ outside, i.e.,
\[\psi :\mathbb{R}^2\times[0,2\pi] \rightarrow\{1,0\}, \quad
\psi (\hat y,\theta)=\left\{
\begin{aligned}
&1,&\quad(\hat y,\theta)\in\Omega(\theta),\\
&0,&\quad(\hat y,\theta)\notin\Omega(\theta).\\
\end{aligned}
\right.\]
An illustration of $\psi$ is shown in Figure \ref{fig:2}. Therefore, the dose distribution function $z(x)$ can also be expressed as
\be \label{eq:z} z(x)= \int_0^{2\pi} \int_{\mathbb{R}^2}
D(x,\hat y,\theta)s(\theta) \psi (\hat y,\theta)~\mathrm{d}\hat y
~\mathrm{d}\theta. \ee

The goal of treatment planning for VMAT is to determine the optimal aperture shapes and beam intensities so that the final dose distribution function $z(x)$ is as close to a prescribed treatment plan as possible. Let $S_r,r=1,\ldots,n_S $, denote different structures in the patient's body which are classified into three categories. The healthy tissues and organs that are very close to the tumors are called critical structures and labeled as the first category \textbf{$I_1$}. The cancerous tissues to be eliminated are called target structures and are assigned to \textbf{$I_2$}. The remaining tissue types are called remainder structures and will be included in the third category \textbf{$I_3$}. Let $m_r$ be the maximum dosage that is allowed for the critical and
remainder structures, and be the necessary dosage to kill the cancer cells for target tissues, respectively.

We proceed to the definition of an energy functional that measures how far the dose distribution $z(x)$ is from the prescribed treatment plan.  For critical and remainder structures, there is no penalty on placing less radiation than the maximum allowable amount. There are only penalties on exceeding the
dose limit. On the other hand, for the target structures, there is a penalty as long as the dose administered is not equal to the dose needed. Allowing arbitrary high dose on tumours may seem acceptable at first glance, while it may cause unexpected damage on healthy tissues. For a given dose distribution $z(x)$, the aforementioned penalty functional for each structure $S_r$ takes the form
\[F_r(z)=\int_{S_r} P_r(z(x))~ \mathrm{d}x,\]
where
\begin{equation}\label{eq:1}
P_r(z(x))=\left\{
\begin{aligned}
&\beta_r(\max\{0,z(x)-m_r\})^2,&&r \in \textbf{$I_1\cup I_3$},\\
&\alpha _r(\max\{ 0,m_r-z(x)\})^2+\beta_r(\max\{0,z(x)-m_r\})^3,&&r \in
\textbf{$I_2$},
\end{aligned}
\right.
\end{equation}
and  $\alpha_r$ and $\beta_r$ are the penalty parameters for the structure $S_r$. Using a cubic penalty function on the targets when the dose is over the prescribed limit can often lead to more favorable results in our numerical experiments. The penalty parameters are set as $\alpha _r>0$ for $r \in I_2$  and $\beta_r>0$ for all $r$. The parameters $\alpha_r$, $\beta_r$ and $m_r$ should be properly adjusted to yield desired plans for different patients.

Combining the definition of $F_r(z)$ and $z(x)$, we obtain the following total energy functional with respect to the aperture shapes $\psi(\hat y,\theta)$ and beam intensity $s(\theta)$:
\be \label{eq:obj}E(\psi ,s)=\frac{1}{2} \sum _{r=1} ^{n_S} \int _{S_r} P_r \left( \int_0^{2\pi} \int_{\mathbb{R}^2}
D(x,\hat y,\theta)s(\theta) \psi (\hat y,\theta)~\mathrm{d}\hat y
~\mathrm{d}\theta  \right )~\mathrm{d}x.\ee

\subsection{The Constraints}

We next clarify the constraints imposed by the deliverable aperture shapes and beam intensities due to mechanical limitations. For simplicity, we only consider four major constraints as follows.

\begin{enumerate}
\item[(i)] The beam intensity $s(\theta)$ should be bounded, i.e., $s\in [0,M_s]$ for a certain given value $M_s$.

\item[(ii)] Due to the physical restriction on the device, there exists a maximum allowable rate $M$ at which a beam can change its intensity during the rotation. It can be described by the following inequality
 \be \label{eq:con2} \left|\frac{d}{d\theta}s(\theta)\right|
 \leq M\ee for all $\theta \in [0,2 \pi]$.

\item[(iii)] An aperture is formed when the two sets of MLC leaves move back and forth. Hence, the deliverable aperture shapes must satisfy a  directional-convexity requirement. Let $w$ be an unit vector in the direction that the MLC leaves are oriented. The physical constraint requires that $\psi (\hat y+\zeta \omega,\theta)=1$ for any $\zeta$ between 0 and $\tau$ if $\psi (\hat y,\theta)=1$ and $\psi (\hat y+\tau \omega,\theta)=1$  for some $\theta $, $\tau\in \mathbb{R}$ and $\hat y$. An illustration is shown by the image on the right in Figure \ref{fig:2}.

\item[(iv)] The MLC leaves can not move faster than a given speed, or in other words, the aperture shapes cannot change too much between two consecutive angles. Let $M_A$ be the maximum speed that the MLC leaves can move. The constraint can be formulated as:
    \be \label{eq:con4}\frac{\left|(\nabla\ \psi \times \omega^{\perp})\cdot \omega\right|}{\left|\nabla\psi \times \omega^{\perp}\right|}\leq M_A,\ee
    where $\omega^{\perp}$ is a unit vector orthogonal to the MLC plane. A simpler discretized formulation of \eqref{eq:con4} will be used later.
\end{enumerate}

\subsection{The Discretized Model}

\begin{figure}[htp]
\centering
\subfigure{
\begin{minipage}[t]{0.45\linewidth}
\centering
\includegraphics[width=\textwidth]{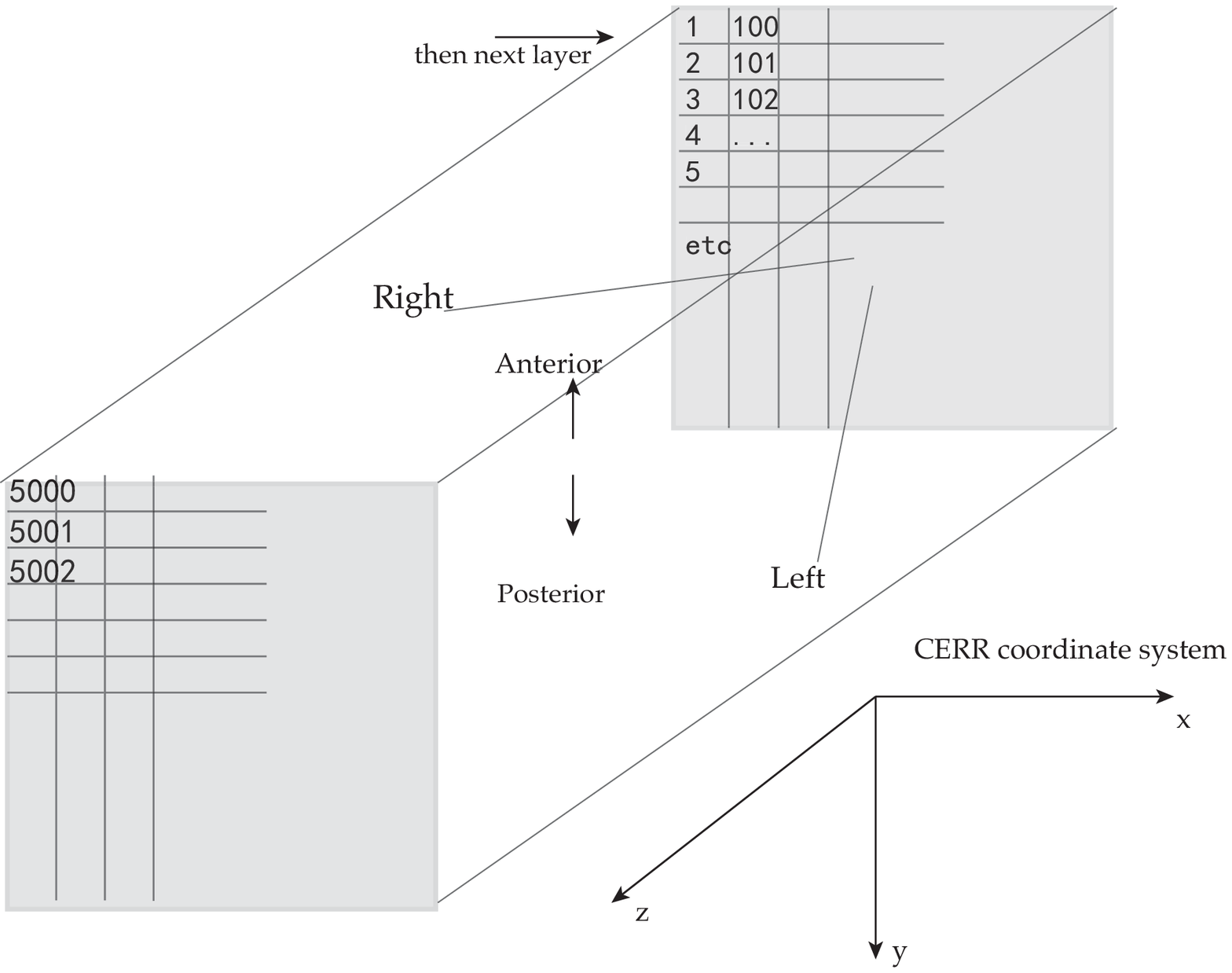}
\end{minipage}
\begin{minipage}[t]{0.45\linewidth}
\centering
\includegraphics[width=0.95\textwidth]{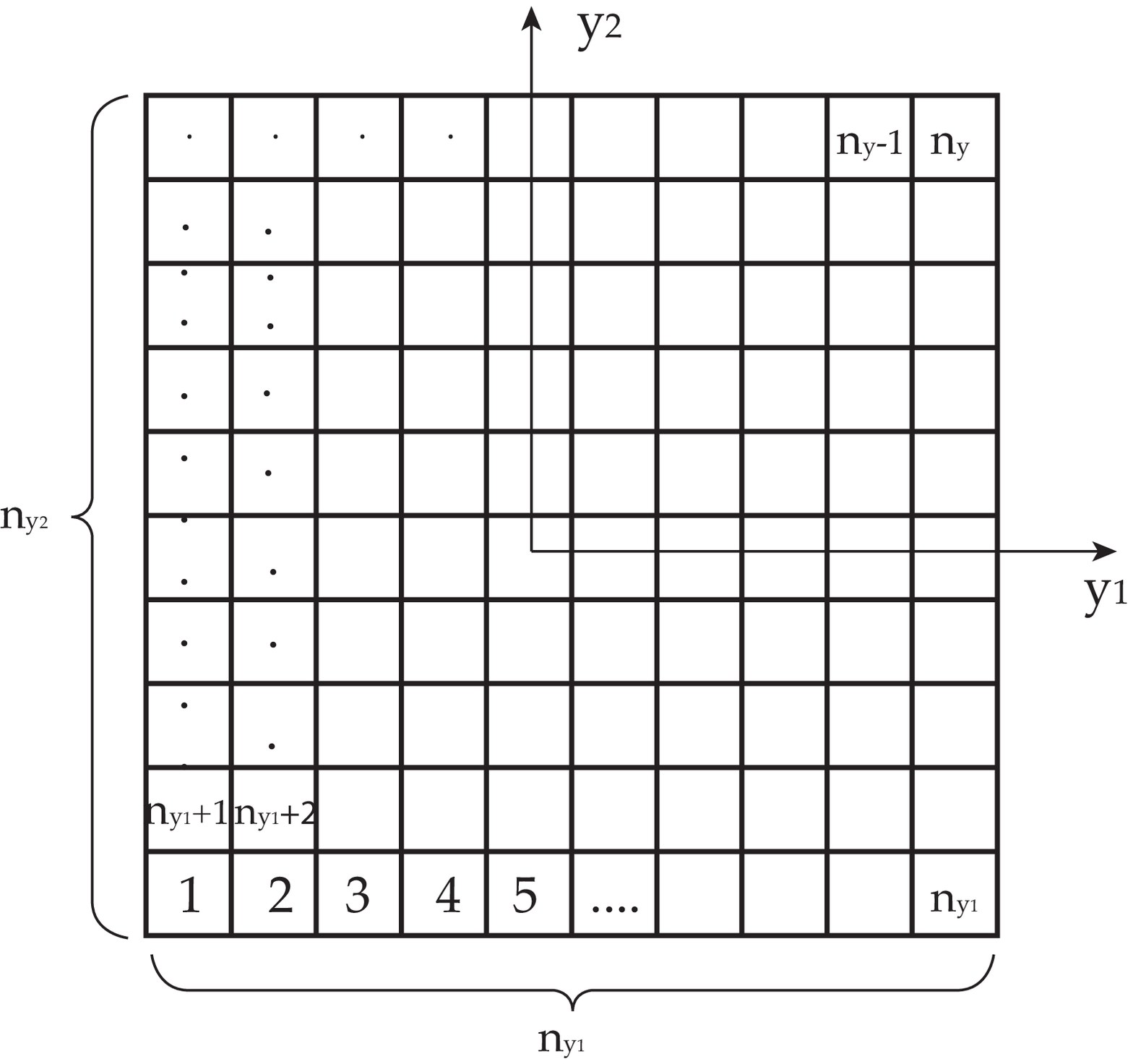}
\end{minipage}}
\caption{Left: An illustration of the voxel indices pattern
of the patient's body of interest, the patient's orientation and the CERR coordinate system \cite{Data2014}.
Right:  An illustration of the indices pattern of the discretized MLC plane.}\label{fig:3}
\end{figure}

Now, we describe how the energy functional \eqref{eq:obj} and the four constraints are discreitized. Recall that $S_r$, $r=1\ldots,n_S$, are domains in $\mathbb{R}^3$ that enclose different patient structures. Let $S\subset\mathbb{R}^3$ be a cubical computation domain such that $\cup_r S_r\subseteq S$. We discretize $S$ using a regular grid that divides $S$ into $n_x$ voxels which leads to a set of voxels takes the form $\{x_i\in S: i=1,2,\ldots,n_x\}$ (see the image on the left in Figure \ref{fig:3}). Such discretization naturally leads to a discretization of $S_r$ as well. With an abuse of terminology, we still denote the discrete voxel set of the $r$-th structure as $S_r$ for simplicity. Similarly, we discretize the MLC plane into $n_y=n_{y_1} n_{y_2}$ regular grids, where $n_{y_1}$ is the number of grids per row in $y_1$ direction, and $n_{y_2}$ is the number of grids per line in $y_2$ direction. Let $\hat y_j \in\mathbb{R}^2 (j=1,2,\ldots,n_y)$ denote the grid point with index $j$ in the MLC plane (see the image on the right in Figure \ref{fig:3}). The range-of-rotation of the gantry is discretized into $n_{\theta}$ angles, where $\theta_k~(k=1,2,\ldots,n_{\theta})$ is the $k$-th gantry angle. For convenience, we use $z_i$, $s_k$, $\psi_{jk}$ and $D_{ijk}$ to denote $z(x_i)$, $s(\theta_k)$, $\psi (\hat y_j,\theta_k)$ and $D(x_i,\hat y_j,s_k)$, respectively. The index $i$ is reserved for the voxel index in patient domain $S$, $j$ is the index of the grid points in aperture domain, and $k$ is the index of gantry angles. Now, the discrete form of $z(x)$ can be written as
\be \label{eq:z1}
z_i=\sum\limits_{k=1}^{n_\theta}\sum\limits_{j=1}^{n_y}D_{ijk}s_k\psi_{jk}.\ee
The integration $F_r(z)$ on a structure $S_r$ can be approximated by a summation of $P_r(z_i)$ over all voxels $x_i$ in the structure $S_r$. Consequently, the total energy functional takes the following discrete form:
\[E(\psi,s)=\frac12\sum\limits_{r=1}^{n_S}\sum\limits_{x_i \in S_r} P_r(z_i).\]

The four constraints mentioned in the previous section can be properly discretized as well. The boundedness of beam intensities $s=\{s_k: k=1,\ldots,n_\theta\}$ can be simply written as $0\le s_k\le M_s$ for all $k$. The constraint \eqref{eq:con2} can be discretized as
\[ |s_{k+1}-s_k|\leq M (\theta_{k+1}-\theta_{k}). \]

In order to properly discretize the directional convexity constraint and
\eqref{eq:con4}, we use an alternative representation of the binary level set
function $\psi(\hat y, \theta)$. Since the deliverable aperture shapes have to
satisfy the direction-convexity and the MLC leaves can only move in $y_1$
direction, the aperture can be described by the positions of the first and the
last grids that are opened per row, which are essentially the tips of the left
and right sets of MLC leaves. Let $L_{k,l}$ and $R_{k,l}$ be the indices of the
first (from left to right) and the last open grid (i.e., where the aperture is open) in the $l$-th row of the MLC plane at the $\theta_k$ angle. It is not hard to see that any deliverable shape $\psi$ can be characterized equivalently by
\[(L,R):=\{(L_{k,l},R_{k,l}) \mid  k=1,\ldots,n_{\theta},
 l=1,\ldots,n_{y_2}\}.\]
Then, one can easily rewrite $z_i$ (defined in \eqref{eq:z1}) as
 \be \label{eq:zLR}
 z_i=\sum\limits_{k=1}^{n_\theta}~\sum\limits_{l=1}^{n_{y_2}}~\sum\limits_{j=L_{k,l}}^{R_{k,l}}D_{ijk}s_k.\ee
Then, the direction-convexity on the $l$-th row is simply
 \[(l-1)n_{y_1}\leq L_{k,l}\leq R_{k,l}\leq ln_{y_1},\quad L_{k,l}, R_{k,l}\in
 \mathbb{Z}. \]
The constraint on the leaf speed limit \eqref{eq:con4} can be discretized as
 \[ \frac{|L_{k,l}-L_{k+1,l}|}{\theta_{k+1}-\theta_{k}}\leq M_A,
  \frac{|R_{k,l}-R_{k+1,l}|}{\theta_{k+1}-\theta_{k}}\leq M_A.\]
%\comm{Add explanation on why the second and third inequalities in \eqref{prob:opt} are the discretization of \eqref{eq:con4}? }

Therefore, the discretized VMAT optimization problem can
be written in the following form:
\be\label{prob:opt}\begin{aligned}
\min_{L,R,s}&\quad   E(L,R,s)=\frac12\sum\limits_{r=1}^{n_S}\sum\limits_{x_i \in S_r} P_r(z_i),\\
\st  &\quad z_i=\sum\limits_{k=1}^{n_\theta}~\sum\limits_{l=1}^{n_{y_2}}~\sum\limits_{j=L_{k,l}}^{R_{k,l}}D_{ijk}s_k,
\quad i=1,\ldots,n_x,\\
&\quad  \frac{|L_{k,l}-L_{k+1,l}|}{\theta_{k+1}-\theta_{k}}\leq M_A,
\quad k=1,\ldots,n_{\theta}-1, l=1,\ldots,n_{y_2},\\
&\quad  \frac{|R_{k,l}-R_{k+1,l}|}{\theta_{k+1}-\theta_{k}}\leq M_A,
\quad k=1,\ldots,n_{\theta}-1, l=1,\ldots,n_{y_2},\\
&\quad  (l-1)n_{y_1}\leq L_{k,l}\leq R_{k,l}\leq ln_{y_1}, \quad  k=1,\ldots,n_{\theta}, l=1,\ldots,n_{y_2},\\
&\quad   L_{k,l}, R_{k,l}\in \{1,2,\ldots,n_{y_1}\},\quad  k=1,\ldots,n_{\theta}, l=1,\ldots,n_{y_2},\\
&\quad  |s_{k+1}-s_k|\leq M (\theta_{k+1}-\theta_{k}),\quad k=1,\ldots,n_{\theta}-1,\\
&\quad s_k\in[0,M_s], \quad k=1,\ldots,n_{\theta}.\\
\end{aligned}\ee
The model \eqref{prob:opt} is a mixed-integer nonlinear and nonconvex optimization problem with respect to the
aperture shapes $(L,R)$, which are integer variables, and the beam intensities $s=\{s_k: k=1,\ldots,n_\theta\}$, which are continuous variables. Note that $z$ is an intermediate variable introduced for convenience.

\section{Algorithms For VMAT Optimization} \label{sec:alg}

Since solving the VMAT optimization problem \eqref{prob:opt} simultaneously with respect to both the aperture $(L,R)$ and the beam intensity $s$ can be difficult, we adopt a simple alternating minimization framework. At each iteration, the aperture shapes are updated by fixing the beam intensity; then the beam intensity is updated while aperture shapes are fixed. This procedure is repeated until convergence. To further reduce computation cost and improve quality of the results, certain randomized strategies are adopted.

\subsection{A Greedy Sampling Algorithm For Finding Aperture Shapes}

For fixed beam intensities $s=\{s_k\}$, the optimization problem with respect to
the aperture shapes is:
\be\label{prob:opt-ap}\begin{aligned}
\min_{L,R}&\quad   E(L,R,s)=\frac12\sum\limits_{r=1}^{n_S}\sum\limits_{x_i \in S_r} P_r(z_i),\\
\st  &\quad
z_i=\sum\limits_{k=1}^{n_\theta}~\sum\limits_{l=1}^{n_{y_2}}~\sum\limits_{j=L_{k,l}}^{R_{k,l}}D_{ijk}s_k,\quad i=1,\ldots,n_x, \\
&\quad  \frac{|L_{k,l}-L_{k+1,l}|}{\theta_{k+1}-\theta_{k}}\leq M_A,
\quad k=1,\ldots,n_{\theta}-1, l=1,\ldots,n_{y_2},\\
&\quad  \frac{|R_{k,l}-R_{k+1,l}|}{\theta_{k+1}-\theta_{k}}\leq M_A,
\quad k=1,\ldots,n_{\theta}-1, l=1,\ldots,n_{y_2},\\
&\quad  (l-1)n_{y_1}\leq L_{k,l}\leq R_{k,l}\leq ln_{y_1}, \quad  k=1,\ldots,n_{\theta}, l=1,\ldots,n_{y_2},\\
&\quad   L_{k,l}, R_{k,l}\in \{1,2,\ldots,n_{y_1}\},\quad  k=1,\ldots,n_{\theta}, l=1,\ldots,n_{y_2}.
\end{aligned}\ee
Since solving the above nonconvex mixed-integer programming directly is challenging, we develop a greedy algorithm inspired by the heuristic strategy proposed by \cite{Dong2012}.

Due to the constraints on the speed limit of the leaves and the directional-convexity of the aperture shapes, the leaf pairs $(L,R)=\{(L_{k,l},R_{k,l})\}$ cannot change too fast. The basic idea of our algorithm is to only update the values of leaf pair $(L_{\hat k,\hat l},R_{\hat k,\hat l})$ one at a time. Given an angle $\theta_{\hat k}$ and the amount of movement $\delta=(\delta_1,\delta_2)$ of the left and right tips of the $\hat l$-th row of the MLC leaves, the
next position $(\widehat L, \widehat R)$ can be described by
\be \label{eq:posn} \widehat L_{ij} =\begin{cases} L_{\hat k,\hat l} + \delta_1, & \mbox{
  if } i=\hat k \mbox{ and } j = \hat l,  \\
  L_{ij}, & \mbox{otherwise,}\end{cases} \quad\widehat R_{ij} =\begin{cases}
    R_{\hat k,\hat l} + \delta_2, & \mbox{
  if } i=\hat k \mbox{ and } j = \hat l,  \\
  R_{ij}, & \mbox{otherwise.}\end{cases}\ee
There are two types of elementary movements of $(\delta_1,\delta_2)$, i.e.,
\be \label{eq:delta-move} \Delta_{\hat k, \hat l}^{L,c} =  \{(a,b) \mid b = 0, \quad  ~ a \in
\Z_c\}, \quad
\Delta_{\hat k, \hat l}^{R,c}= \{(a,b) \mid ~a=0, \quad  b \in \Z_c\}.\ee
where $\Z_c = \{ z| -c \le z \le c, z \in \Z\}$. A value $\delta\in \Delta_{\hat k,\hat l}^{L,c}$ means that the left tip of the leaf can move in a small neighborhood of $L_{\hat k,\hat l}$ with respect to the first index with the right tip fixed, and similarly for $\delta\in \Delta_{\hat k,\hat l}^{R,c}$.
It is easy to verify that these two types of movements can be combined to generate general movements. Hence, we focus on the movements in $\Delta_{\hat k, \hat l}^{L,c}$ and  $\Delta_{\hat k, \hat l}^{R,c}$. Our aperture algorithm is to move the leaf pair row by row and the changes are accumulated until the optimal shapes are obtained. In fact, the greedy algorithm proposed in \cite{Dong2012} is a special case of \eqref{eq:delta-move} with $c=1$, while our method has more flexibility in changing the aperture shapes by searching in a larger region defined by  \eqref{eq:delta-move}.

After each movement \eqref{eq:posn}, the change of the energy function can be relatively easily calculated because the update only modifies the dosages on a small portion of the total voxels when $c$ is small. For example, if $\delta_1 > 0$ and $\delta_2=0$, we have
\[ \hat z_i = z_i - \sum\limits_{j=L_{\hat k,\hat l}}^{L_{\hat k,\hat l} +
\delta_1 - 1}D_{ij\hat k}s_{\hat k},\]
which yields
\begin{equation}\label{eq:2}
\Delta E_{\hat k,\hat l}^{\delta}= E(\widehat L, \widehat R,s)- E(L, R,s)
=\frac12\sum\limits_{r=1}^{n_S}\sum\limits_{x_i \in S_r, \hat z_i \neq z_i} \left( P_r(\hat z_i) - P_r(z_i)\right).
\end{equation}
Similar relationships hold for other values of $\delta$. When $c$ is set to $1$ in \eqref{eq:delta-move}, the number of operations needed to determine $\Delta E_{\hat k,\hat l}$  is of the order of the number of $i$ such that $D_{ij\hat k}\neq 0$. Consequently, the best possible local modification of the left tip is a movement $\delta \in \Delta_{\hat k,\hat l}^{L,c}$ such that $\Delta E_{\hat k,\hat l}^\delta \le 0$ and the resulting position $(\hat L, \hat R)$  represents a deliverable aperture shape.  The best one from the right side of the leaf can be found in the same fashion.  For convenience, we denote these two movements as follows:
 \bea \delta_{\hat k,\hat l}^L &=& \arg \min_{\delta}\left\{\Delta E_{\hat k,\hat
 l}^\delta \mid \delta \in \Delta_{\hat k,\hat l}^{L,c}, \quad \Delta E_{\hat k,\hat
 l}^\delta \le 0, \quad  (\hat L, \hat R) \mbox{
 is feasible}\right\},\\
 \delta_{\hat k,\hat l}^R &=& \arg \min_{\delta}\left\{\Delta E_{\hat k,\hat
 l}^\delta \mid \delta \in \Delta_{\hat k,\hat l}^{R,c}, \quad \Delta E_{\hat k,\hat
 l}^\delta \le 0, \quad  (\hat L, \hat R) \mbox{
 is feasible}\right\}. \eea
Note that the values of $\delta_{\hat k,\hat l}^L$ and  $\delta_{\hat k,\hat l}^R$ might be zero if there is no better position than $(L,R)$.

We next present the rules on selecting elements $(\hat k, \hat l)$ to update sequentially. In order to save on computation time, we create a heap at the beginning and only update elements in the heap based on the following strategies. Denote the collection of all feasible movements that lead to a strict decreasing of the energy function from the current aperture shape by
 \be \label{eq:gamma} \Gamma:= \left\{ (\hat k, \hat l, \delta) \mid \Delta E_{\hat k,\hat
 l}^\delta <0, \hat k=1,\ldots,n_{\theta}, \hat
l=1,\ldots,n_{y_2}, \delta \in \left\{ \delta_{\hat k,\hat l}^{L},\delta_{\hat
k,\hat l}^{R} \right\}\right\}. \ee
Then we sort all elements in $\Gamma$ so that the corresponding values  $\Delta
E_{\hat k,\hat l}^\delta$ are in an ascending order and we still denote the
sorted list as $\Gamma$. Obviously, the first element in $\Gamma$ has the
smallest $\Delta E_{\hat k,\hat l}^\delta$ and it is the best local refinement
in $\Gamma$. We update the aperture shape using this element and delete it from
the heap $\Gamma$ afterwards. Since this change may violate the feasibility of
other values in $\Gamma$, we recompute $ \Delta E_{\hat k,\hat l}^\delta$ for
the next $\min(\Upsilon, |\Gamma_\Pi|)$ elements in line in $\Gamma$ and fix
their ordering. The value of $\Upsilon$ is usually set to $5$ to $10$ for saving
the computational cost. This procedure is repeated and iterated until $\Gamma$ is empty.

The major computational cost of the above greedy procedure is calculating
$\Delta E_{\hat k,\hat l}^\delta$ repeatedly for all $\hat
k=1,\ldots,n_{\theta}$ and $\hat l=1,\ldots,n_{y_2}$. However, since only one
row of the aperture shapes for a specific value of $\delta$ is updated at each
iteration,  computing all possible  $\Delta E_{\hat k,\hat l}^\delta$ in the
heap $\Gamma$ is definitely not the best strategy. If we randomly take a small
number of leaf pairs, the computational cost can be significantly reduced. The
randomness in sampling may also prevent the algorithm from being trapped in a
local minimum at an early stage. To be more precise, given a sampling ratio
$0<\kappa_{\Pi}\le 1$,  we take $\kappa_\Pi$ uniformly sampled elements
  from $\Gamma$ as
\be \label{eq:gamma1} \Gamma_{\Pi}:= \left\{ (\hat k, \hat l, \delta) \mid
(\hat k, \hat l, \delta) \in \Gamma \right\} \mbox{ and } |\Gamma_\Pi|=
\kappa_{\Pi} |\Gamma|. \ee
Then the randomized greedy algorithm only updates elements in the set
$\Gamma_\Pi$.
In our experiments, the amount of change of aperture shapes decreases quickly for each iteration. Therefore, we start with fully sampling $\kappa_{\Pi}=1$ and gradually decrease the sampling rate at each iteration.

% To be more precise, given a sampling ratio $0<\kappa_{\Pi}\le 1$,  we define
% the uniformly sampled index set $\Pi$ as \be \label{eq:sPi} \Pi=\{ (k,l) \mid
% 1\le k \le n_{\theta}, 1 \le l \le n_{y_2}\} \quad\mbox{and}\quad
% |\Pi|=\kappa_{\Pi} n_{\theta}n_{y_2}. \ee Then, the randomized greedy
% algorithm only updates elements in the set $\Pi$: \be \label{eq:gamma1}
% \Gamma_{\Pi}:= \left\{ (\hat L, \hat R) \mid \Delta E_{\hat k,\hat l}^\delta <0, (\hat k, \hat l)\in \Pi, \delta \in \left\{ \delta_{\hat k,\hat l}^{L},\delta_{\hat k,\hat l}^{R} \right\}\right\}. \ee In our experiments, the amount of change of aperture shapes decreases quickly for each iteration. Therefore, we start with fully sampling $\kappa_{\Pi}=1$ and gradually decrease the sampling rate at each iteration.

We outline our aperture optimization algorithm as follows.

\begin{algorithm2e}[H]\caption{A greedy algorithm for aperture optimization}
\label{alg:ap1}
Input an initial aperture $(L,R)$, beam intensities $s$ and a sampling ratio
$\kappa_{\Pi}$. Set $c, \Upsilon\ge 1$. \\% and $n_u=0$. \\
%Take a random sample set $\Pi$ defined by \eqref{eq:sPi} with the sampling ratio $\kappa_{\Pi}$. \\
Compute an initial heap $\Gamma_{\Pi}$ defined by \eqref{eq:gamma1} of all feasible movements based on
\eqref{eq:gamma} and reorder all elements
in $\Gamma_{\Pi}$ so that the corresponding values  $\Delta E_{\hat k,\hat
 l}^\delta$ are in an ascending order. \\
%  \label{step}Update the aperture shapes using the first element in
%  $\Gamma_{\Pi}$ and delete it from $\Gamma_{\Pi}$.\\
%  Recompute $ \Delta E_{\hat k,\hat l}^\delta$ for all elements in $\Gamma_{\Pi}$ with fixed the ordering.  \\
 Take the first element $(\hat k, \hat l, \delta)$ in $\Gamma_\Pi$.
 Update the aperture shapes using $(\hat k, \hat l, \delta)$ if  $\Delta E_{\hat
 k,\hat l}^\delta<0$  and it represents a
deliverable aperture shape. \\
  Delete $(\hat k, \hat l, \delta)$  from $\Gamma_{\Pi}$. Recompute $ \Delta
  E_{\hat k,\hat l}^\delta$ for the next $\min(\Upsilon, |\Gamma_\Pi|)$ elements in
 the new $\Gamma_{\Pi}$ and fix the ordering.  \\
%  \label{step} \textcolor{blue}{Recompute $ \Delta E_{\hat k,\hat l}^\delta$ for the first element in $\Gamma_{\Pi}$, if  $\Delta E_{\hat k,\hat l}^\delta<0$ and this movement will not violate the constraints, update the aperture shapes using this element. \\ Delete the first element from $\Gamma_{\Pi}$}.\\
 If $\Gamma_{\Pi}$ is non-empty, go to step 3, else return the new aperture shapes $(L,R)$.% and $n_u$.
\end{algorithm2e}

\subsection{Projected Gradient Methods for Beam Intensity Optimization}

For fixed aperture shapes at different angles $(L,R)$, the optimization problem with respect to the beam intensities $s$ is:
\be\label{prob:opt-sk}\begin{aligned}
\min_{s}&\quad   E(L,R,s)=\frac12\sum\limits_{r=1}^{n_S}\sum\limits_{x_i \in S_r} P_r(z_i),\\
\st  &\quad
z_i=\sum\limits_{k=1}^{n_\theta}~\sum\limits_{l=1}^{n_{y_2}}~\sum\limits_{j=L_{k,l}}^{R_{k,l}}D_{ijk}s_k,\quad i=1,\ldots,n_x, \\
&\quad  |s_{k+1}-s_k|\leq M (\theta_{k+1}-\theta_{k}),\quad k=1,\ldots,n_{\theta}-1,\\
&\quad s_k\in[0,M_s], \quad k=1,\ldots,n_{\theta}.
\end{aligned}\ee
The main difficulty in solving \eqref{prob:opt-sk} is the nonlinearity of the objective function $E(L,R,s)$. Since the constraints are relatively simple, we
apply the widely used projected gradient method to solve \eqref{prob:opt-sk}.

The partial derivative of the energy function with respect to $s_k$ is given as follows:
\[\begin{aligned}
\frac{\partial E(L, R ,s)}{\partial s_k}
&= \frac12\sum\limits_{r=1}^{n_S} \sum_{x_i \in S_r} P'_r(z_{i})\frac{\partial z_i}{\partial s_k}
= \frac12\sum\limits_{r=1}^{n_S} \sum_{x_i \in S_r} P'_r (z_i)d_{ik},\\
\end{aligned}\]
where
$d_{ik}=\sum\limits_{l=1}^{n_{y_2}}~\sum\limits_{j=L_{k,l}}^{R_{k,l}}D_{ijk}$,
and
\[
P'_r(z)=\left\{
\begin{aligned}
&2\beta_r(\max\{0,z-m_r\}),
&&\quad  ~~r \in \textbf{$I_1\cup I_3$},\\
&2\alpha _r(\max\{ 0,m_r-z\})+3\beta_r(\max\{0,z-m_r\})^2,
&& \quad  ~r \in \textbf{$I_2$}.\\
\end{aligned}
\right.
\]

Denote $s^{(q)}$ as the approximated optimal intensity values at the $q$-th iteration, and $\nabla_s E^{(q)}:=\nabla_s E(L,R,s^{(q)})$. Given $s^{(q)}$ and a step size $\tau^{(q)}$, the projected gradient method first approximates the energy function by linearizing it with respect to $s$ and adding a proximal term as
\beaa  &&E(L,R,s^{(q)}) + \left(\nabla_s E^{(q)}\right)^\top (s-s^{(q)}) +
  \frac{1}{2\tau^{(q)}} \|s-s^{(q)}\|^2 \\
  &=& \frac{1}{2\tau^{(q)}} \|s- \left( s^{(q)} -
  \tau ^{(q)} \nabla_s E^{(q)}\right)\|^2 + \mbox{ constant},\eeaa
then computes a new trial point $s(\tau^{(q)})$ as the optimal solution of
\be\label{prob:opt-sk-grad}\begin{aligned}
  %s(\tau^{(q)})=\arg
  \min_{s}&\quad     \frac{1}{2} \left\|s- \left( s^{(q)} -
  \tau ^{(q)} \nabla_s E^{(q)}\right) \right\|_2^2,\\
\st &\quad  |s_{k+1}-s_k|\leq M (\theta_{k+1}-\theta_{k}),\quad k=1,\ldots,n_{\theta}-1,\\
&\quad s_k\in[0,M_s], \quad k=1,\ldots,n_{\theta}.
\end{aligned}\ee
The subproblem \eqref{prob:opt-sk-grad} is a standard quadratic problem whose objective function and constraints are simple. Since the length $n_{\theta}$ is
often relatively small, \eqref{prob:opt-sk-grad} can be solved efficiently by commercial solvers such as Mosek.

Another key algorithmic issue is the determination of a suitable step size $\tau^{(q)}$. Instead of using the classical Armijo-Wolfe based monotone line search, we apply the nonmonotone line search determined by the BB formula, which is proposed by Barzilai and Borwein \cite{BB1988}  in 1988. We have found that the BB method is more efficient than the monotone line search for our problem. At iteration $q$, the step size is computed by
\begin{equation}\label{eq:6}
\tau_{BB_1}^{(q)}=
\frac{\left(\nu^{(q-1)}\right)^Ty^{(q-1)}}{\left(y^{(q-1)}\right)^Ty^{(q-1)}},\quad
\tau_{BB_2}^{(q)}=\frac{\left(\nu^{(q-1)}\right)^T\nu^{(q-1)}}{\left(\nu^{(q-1)}\right)^Ty^{(q-1)}},
\end{equation}
where $y^{(q-1)}=\nabla_s E^{(q)}-\nabla_s E^{(q-1)}$ and $\nu^{(q-1)}= s^{(q)}- s^{(q-1)}$. In order to guarantee convergence, the final value of $\tau^{(q)}$ is a fraction of $\tau_{BB_1}^{(q)}$ and $\tau_{BB_2}^{(q)}$ determined by a nonmonotone search condition in \cite{ZhangHager2004}. Let  $C^{(0)}=E(L,R,s^{(0)})$, $ V^{(q+1)} = \eta V^{(q)} +1$ and  $V^{(0)}=1$. The new points
are generated iteratively in the form  $s^{(q+1)}= s(\tau^{(q)} )$, where $s(\tau^{(q)} )$ is the solution of subproblem \eqref{prob:opt-sk-grad}, $\tau^{(q)} =\sigma^t \tau_{BB_1}^{(q)}$ or  $\tau^{(q)} = \sigma^t\tau_{BB_2}^{(q)}$, $\sigma \in (0,1)$ and $t$ is the smallest nonnegative integer satisfying
\begin{equation} \label{eq:NMLS-Armijo}
E(L,R, s^{(q+1)}) \le C^{(q)} - \frac{\gamma}{2} \| s^{(q+1)}- s^{(q)}\|_F^2,
\end{equation}
where $\gamma>0$, each reference value $C^{(q+1)}$ is taken to be the convex combination of  $C^{(q)}$ and $E(L,R, s^{(q+1)})$ as $C^{(q+1)} = (\eta V^{(q)}
C^{(q)} + E(L,R, s^{(q+1)}))/V^{(q+1)}$.  Since we observe that the BB step size $\tau_{BB1}^{(q)}$ or $\tau_{BB2}^{(q)}$ is often sufficient for
\eqref{eq:NMLS-Armijo} to hold in our numerical experiments, we set the initial value of $\tau^{(q)}$ as
 \begin{equation}\label{eq:8}
 \tau^{(q)}=\left\{
 \begin{aligned}
&\tau_{BB1}^{(q)},&&\quad  ~q~\text{is~odd},\\
&\tau_{BB2}^{(q)},&&\quad  ~q~\text{is~even}.\\
 \end{aligned}
 \right.
\end{equation}

Our method on finding the beam intensities is presented in Algorithm \ref{alg:NPG}. We terminate the algorithm when the change of the energy function is small comparing to the initial value $E(L,R,s^{(0)})$, i.e.,
 \[ |E(L,R,s^{(q+1)})-E(L,R,s^{(q)})| \le \epsilon_b |E(L,R,s^{(0)})|,\]
where $\epsilon_b$ is a small positive number.

\begin{algorithm2e}[H]\caption{A nonmonotone projected gradient method for beam
  intensities}
\label{alg:NPG}%\LinesNumberedHidden
Given $(L,R)$ and $s^{(0)}$, set $\epsilon_b, \tau^{(0)}>0$,  $ \gamma, \sigma, \eta \in (0,1)$,
$q=0$, $V^{(0)}=1$, $C^{(0)}=E(L,R,s^{(0)})$. \\
\While{ convergence is not met }{
Find the smallest nonnegative integer $t$ such that the optimal solution
$s(\tau^{(q)})$ of \eqref{prob:opt-sk-grad}  satisfies
condition \eqref{eq:NMLS-Armijo}.
Set  $s^{(q+1)}\gets s(\tau^{(q)})$. \\
Compute the step size $\tau_{BB1}^{(q)}$ or $\tau_{BB2}^{(q)}$ according to
\eqref{eq:6} and \eqref{eq:8}.\\
Update {$V^{(q+1)}\gets\eta V^{(q)}  +1$ and $C^{(q+1)} = (\eta V^{(q)}
C^{(q)} + E(L,R, s^{(q+1)}))/V^{(q+1)}$ }.\\
  $q\gets q+1$.
}
\end{algorithm2e}

%\subsection{Randomized algorithm for VMAT}

\subsection{Incremental Importance Sampling in Computing the Energy
Function}

Since the number $n_x$ of voxels can easily be as large as several millions, calculating the dose distribution $z$ and evaluating the energy function
$E(L,R,s)$ are computationally expensive in Algorithms \ref{alg:ap1} and \ref{alg:NPG}.  However, the similarity among the same type of tissues and organs often leads to a large amount of uniformity in the measurements. This indicates that a full evaluation of $E(L,R,s)$ and its gradient $\nabla_s E(L,R,s)$  may not be necessary to make sufficient progress in solving \eqref{prob:opt}. This observation motivates our incremental importance sampling strategy, which only evaluates the objective function $E(L,R,s)$ with respect to a small number of carefully selected voxels. Comparing to the classical incremental randomized algorithms, our selection of the voxels is based on the importance of the voxels measured by certain rules. The cost of incremental importance sampling methods is proportional to the full size methods. In order to eventually achieve a high accuracy, the sampling ratio is increased gradually as the number of iteration increases.

We now introduce the details on how the voxels are chosen. Generally speaking,
remainder structures are not as important as critical or target structures in
the treatment planning.  It is reasonable to treat these voxels in the remainder
structures as equally important. Also, eliminating the target structures and
protecting healthy tissues and organs are equally important. For voxels in
targets and tissues/organs, we define two different importance metrics
respectively. For target structures, the penalty function $P_r(z_i)$ defined in
\eqref{eq:1} is a reasonably good metric since it measures how much each voxel
contributes to the total energy. This function is nonzero on almost every voxels in these structures.
 For critical structures, using $P_r(z_i)$ may not be the best option because voxels in healthy tissues/organs that receive dosage below $m_r$ have a penalty
 value $P_r(z_i)=0$ and these voxels may be the majority of voxels of critical structures. The voxels with the same penalty value will be considered
 equally important and sampled with the same probability. Certainly, those who receive
 more dosage are in a more dangerous position and we should pay more attention
 on them.  Hence, the ratio between the dosage $z_i$ and the maximum
 dosage can be a good indicator of importance since it provides a larger weight
 to  the voxels  with a higher dosage and $P_r(z_i)=0$. Note that a portion of the
 voxels may still receive zero dosage, we further add a small number $h>0$ and assign
a metric $\frac{z_i}{m_r}+h$ to each voxel. 

%On the other hand, because there is no need to protect voxels with a higher dosage in the target structures, the function $\frac{z_i}{m_r}$ is not appropriate for them.

%\textcolor{red}{Why $\frac{z_i}{m_r}$ is better? Because voxels in healthy tissues that receive dosage below $m_r$ all have penalty $P_r(z_i)=0$, and they will be considered equally important if we use $P_r(z_i)$ as their importance. Obviously, it's not reasonable.  Those receiving more dosage are in more danger
%and thus more attention should be paid to them. The importance defined by $\frac{z_i}{m_r}$ can help us attach more importance to voxels with higher dosage.
% However, for targets, it's not the case, since most of the $P_r(z_i)$ are nonzero.  $P_r(z_i)$ actually measures how much each voxel contributes to the total energy, so it is naturally an importance metric. In addition, voxels with higher dosage are not more important any more, because we don't need to protect them. So $\frac{z_i}{m_r}$ is not appropriate for targets.}
%
Let $\widehat S_u=\cup_{r\in I_u} S_r$, $u=1,2,3$. For a specifically given dose distribution $z^*$, the probability of choosing a voxel $x_i \in S_r$ is defined as
\be \label{eq:piz}\pi^{z^*}_{r,i}=\left\{
\begin{aligned}
  & \frac{ \frac{z_i^*}{m_r}+h }{\sum_{x_i \in \widehat S_1} (\frac{z_i^*}{m_r}+h )  },&\quad  \quad  ~~ r\in I_1,\\
  &\frac{P_r(z_i^*)}{\sum_{x_i \in\widehat S_2} P_r(z_i^*) },&\quad  \quad  ~~   r\in I_2, \\
  & \frac{1}{|\widehat S_3|}, & \quad \quad r \in I_3,
\end{aligned}
\right.\ee
where $|\widehat S_u|$ is the total number of voxels in $\widehat S_u$. Let
$0<\kappa_u\le 1$ be the sampling ratio such that $\kappa_u$ is gradually
increased in critical and target structures while $\kappa_u$ is fixed in
remainder structures. The voxels in $\widehat S_u$ are chosen in $c=\kappa_u
|\widehat S_u|$ independent identical trials where in each trial the $i$-th
voxel in $ S_r$ is selected with probability $\pi^{z^*}_{r,i}$, and the
collection of the corresponding indices is denoted by $\B^{z^*}_r$.  In
particular, the samples are chosen without replacement (i.e., every voxel
 occurs only once).

Therefore, the sampled total energy functional is defined as
\begin{equation}\label{eq:10}
  E_{z^*}(L,R,s)=\sum\limits_{u=1}^{3}\frac{1}{\kappa_u}\left(\sum\limits_{r\in I_u}\sum_{i\in \B^{z^*}_r}P_r (z_i)\right).
\end{equation}
Finally, we obtain the randomized version of Algorithms \ref{alg:ap1} and \ref{alg:NPG} by simply replacing $E(L,R,s)$ by the its sampled version  $E_{z^*}(L,R,s)$. Consequently, the computational time is greatly reduced when the sampling ratio $\kappa_u$ is small.

\subsection{Ramdomized Algorithms for VMAT Optimization}

Combining the algorithms for finding the aperture shapes and beam intensities together,  we present our randomized VMAT optimization method in Algorithm \ref{alg:rand}. With an abuse of terminology, the superscripts of beam intensities for the inner iteration $s^{(q)}$ in Algorithm \ref{alg:NPG} and the outer iteration $s^{(n)}$ in Algorithm \ref{alg:rand} are both used whose meaning should be clear from context. We terminate the algorithm when the relative change on the value of the energy function is small, i.e.,
 \[ |E_{z^{(n+1)}}(L^{(n+1)},R^{(n+1)},s^{(n+1)})-E_{z^{(n)}}(L^{(n)},R^{(n)},s^{(n)})| \le \epsilon
 |E_{z^{(n)}}(L^{(n)},R^{(n)},s^{(n)})|,\]
where $\epsilon$ is small positive number. Note that Algorithm \ref{alg:rand}
with $\kappa_{\Pi}=1$ and $\kappa_u =1$ reduces to a deterministic method
without any sampling. %(\textcolor{red}{this is not exactly true since even when $\kappa_u=1$, the probably $\pi^{z^*}_{u,i}$ is not equal to 1 and hence you are still only selecting part of the voxels.}).
%(\textcolor{blue}{I don't think so, because we sample without repetition, and when $\kappa_u =1$, we have to select all the voxels. Since $\pi^{z^*}_{u,i}$ is nonzero for every voxels so we are able to include them all in the end. So it's deterministic}).

%Let $r_1^*$ and $r_2^*$ be two constants. We steady increase the sampling rate in the voxels by $r_2^*$ and
%decrease the sampling rate in the leaf pairs by $r_1^*$. $c_3$ and $c_4$ are two constant. Since randomness is involved,
%the stoping criteria should be changed. We consider to use the mean energy, but the first $c_3$
%iterations are not taken into consideration, because the energy changes dramatically in the first a few iteration.
%And the criteria starts to work after $c_4$ iteration, which can help prevent the algorithm from stopping too early
%and yielding unsatisfactory results. Our Randomized algorithm for VMAT is as follows:

\begin{algorithm2e}[H]\caption{Randomized algorithm for VMAT Optimization}
\label{alg:rand}%\LinesNumberedHidden
Set the initial aperture shapes $(L^{(0)},R^{(0)})$ and beam intensities
$s^{(0)}$. Set the sampling ratio $\kappa_{\Pi}=1, \kappa_u \ge 0$ and $n:=0$.
Set the tolerance value $\epsilon>0$.\\
\While{ convergence is not met }{
Compute $z^{(n)}$ defined by \eqref{eq:zLR}
  using $(L^{(n)}, R^{(n)})$ and $s^{(n)}$.  \\
Compute the sampling probabilities $\pi^{z^*}_{r,i}$ according to
\eqref{eq:piz} with $z^* = z^{(n)}$.  For each type of structure $\widehat S_u$, execute $c=\kappa_u|\widehat S_u|$ independent identical trials where in each trial the $i$-th
voxel in $S_r$ is
chosen with probability $\pi^{z^*}_{r,i}$ and without replacement, and the collection of the corresponding indices
is denoted by $\B^{z^*}_r$.
 \\
Compute the new aperture shapes $(L^{(n+1)}, R^{(n+1)})$ using Algorithm \ref{alg:ap1}
with the sampled energy function \eqref{eq:10}, the aperture shapes $(L^{(n)},
R^{(n)})$, the beam intensities $s^{(n)}$
and the sampling ratio $\kappa_{\Pi}$. \\
Compute the new beam intensities $s^{(n+1)}$ using Algorithm \ref{alg:NPG} with
the sampled energy function \eqref{eq:10} and the aperture shapes $(L^{(n+1)}, R^{(n+1)})$.\\
 Decrease the sampling ratio $\kappa_{\Pi}$ for Algorithm \ref{alg:ap1}.  Increase the sampling ratio $\kappa_u$ for Algorithm \ref{alg:NPG}. \\
  Set $n:=n+1$.
  }
\end{algorithm2e}

The Algorithm \ref{alg:rand} can be further improved in a few ways. For example, we have found that the beam intensity algorithm can be time consuming. The dose distribution changes drastically when the beam intensity changes, which means that one have to recalculate the full dose distribution in every iteration of Algorithm \ref{alg:NPG}. On the other hand, only a small part of the dose distribution changes when the leaves changes in Algorithm \ref{alg:ap1}. Hence, if the change of the aperture shapes is not large, we can skip the step of updating beam intensities and continue refining the aperture shapes.

Although the solutions yielded by Algorithm \ref{alg:rand} will not be exactly the same for every run due to the random sampling, they are still similar to each other. The algorithm is robust since the increase of the sampling ratios on critical and target structures enables the algorithm to eventually take all the voxels into account. In our experiments, the randomized algorithm often returns a solution with better quality and uses less amount of time than the deterministic algorithms.

\section{Numerical results} \label{sec:num}
In this section, we evaluate the performance of our algorithms on two examples, prostate and head-and-neck cancer, from the common optimization for radiation
therapy (CORT) dataset\footnote{Downloadable from \url{http://gigadb.org/dataset/100110}}. More detailed information on the CORT dataset can be found in \cite{Data2014}. We have implemented the method in  \cite{Dong2012} and our Algorithm
\ref{alg:rand} using C.  The algorithm introduced by \cite{Dong2012} shall be referred to as ``GEAltMin'' (Greedy-Euler-flow-Alternating-Minimization). The
deterministic version of Algorithm \ref{alg:rand} with $\kappa_{\Pi}=1$ and $\kappa_u =1$ will be called ``GGAltMin'' (Greedy-Gradient-projetion-Alternating-Minimization).  The randomized version of Algorithm \ref{alg:rand} is called ``RGAltMin'' (Randomized-Greedy-Gradient-projetion-Alternating-Minimization). Our numerical experiments on the comparisons of these three algorithms are preformed on a workstation with two twelve-core Intel Xeon E5-2697 CPUs and 128GB of memory running Ubuntu 12.04. The penalty parameters in the VMAT models are tuned for the best results.

The quality of the computed treatment plans are evaluated through the so-called Dose-Volume histogram (DVH). For a given structure $S_r$, a dose distribution $z(x)$ and the corresponding bound $m_r$, the DVH function for this structure is defined as the percentage of the structure that is radiated at or above the dose level $m_r$, i.e., \begin{equation}\label{eq:11} \text{DVH}(m_r)=\frac{vol(\{x|z(x)\geq m_r\})}{vol(S_r)}, \end{equation} where $vol(S_r)$ denotes the volume of the structure $S_r$.  For a target structure, the optimal DVH curve should remain 100\% before the prescribed dose and drop sharply to 0\% afterwards. For a critical structure, the desired DVH curve should stay 0\% above the safe dose limit for the particular tissue involved.

We first test the prostate case. A structure named PTV\_68  is set as the target, which is a geometric expansion of the gross tumor volume (GTV). For this region, a 73.8 Gy radiation dose is prescribed to eliminate the cancerous growth. For the nearby critical structures of the bladder and rectum, safe dose limits are 23.52 Gy and 32.92 Gy respectively. A summary of computational results is presented in Table \ref{table1}, where ``iteration'' is the total number of the outer iterations for each algorithm, ``$E$'' denotes the value of the energy function and ``runtime'' is the runtime measured in
seconds. Since there are randomness in the run of RGAltMin, its reported results are the average of 20 runs. The table shows that our algorithms take much fewer iterations and much less time to achieve a comparable objective function value. Although the iteration number of RGAltMin is more than that of GGAltMin, the former is still faster than the later because the cost per iteration is cheaper due to the sampling strategies.

\begin{figure}[htp]
\centering
\begin{minipage}[t]{0.48\linewidth}
\centering
\includegraphics[width=\textwidth]{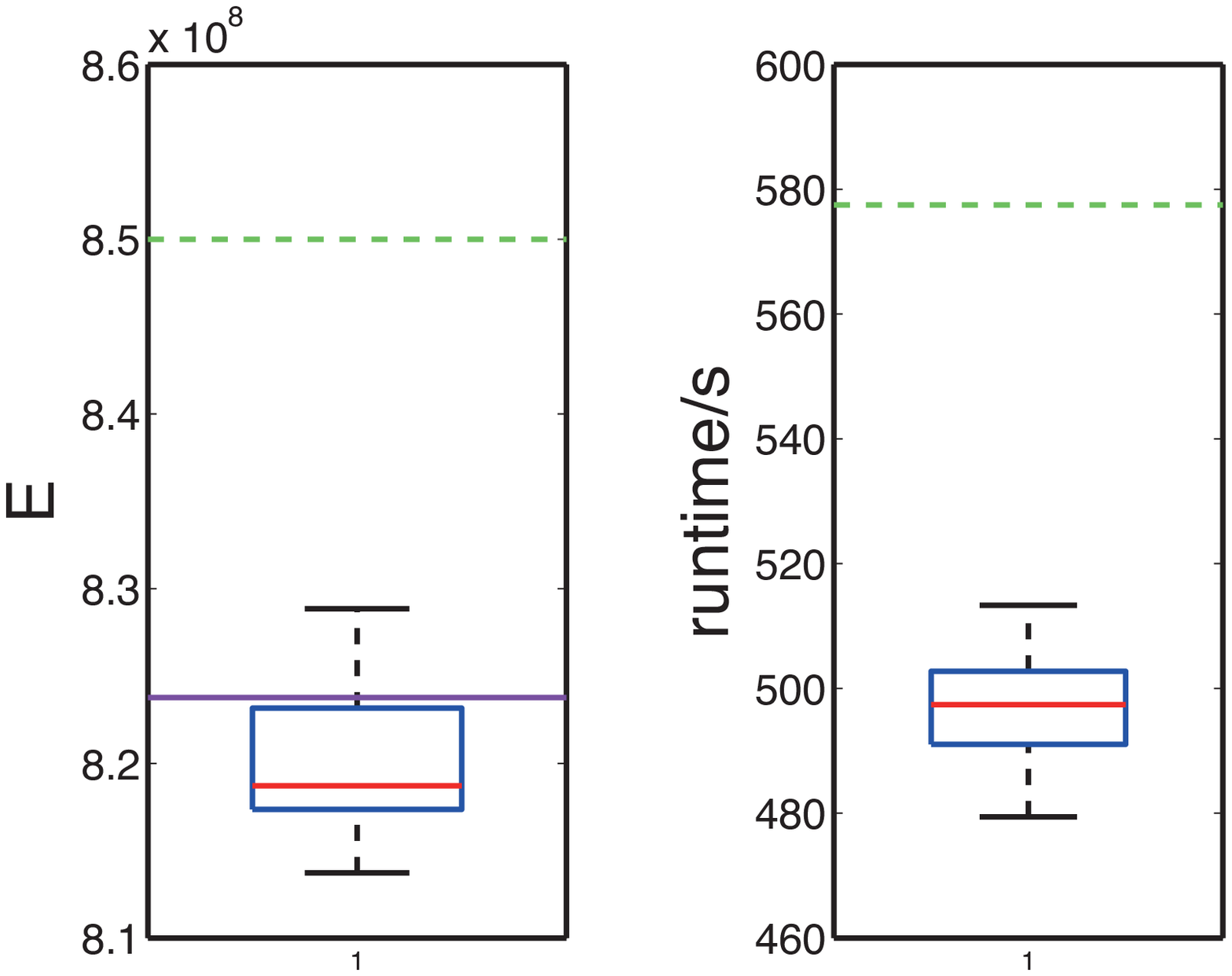}
\end{minipage}
\begin{minipage}[t]{0.48\linewidth}
\centering
\includegraphics[width=\textwidth]{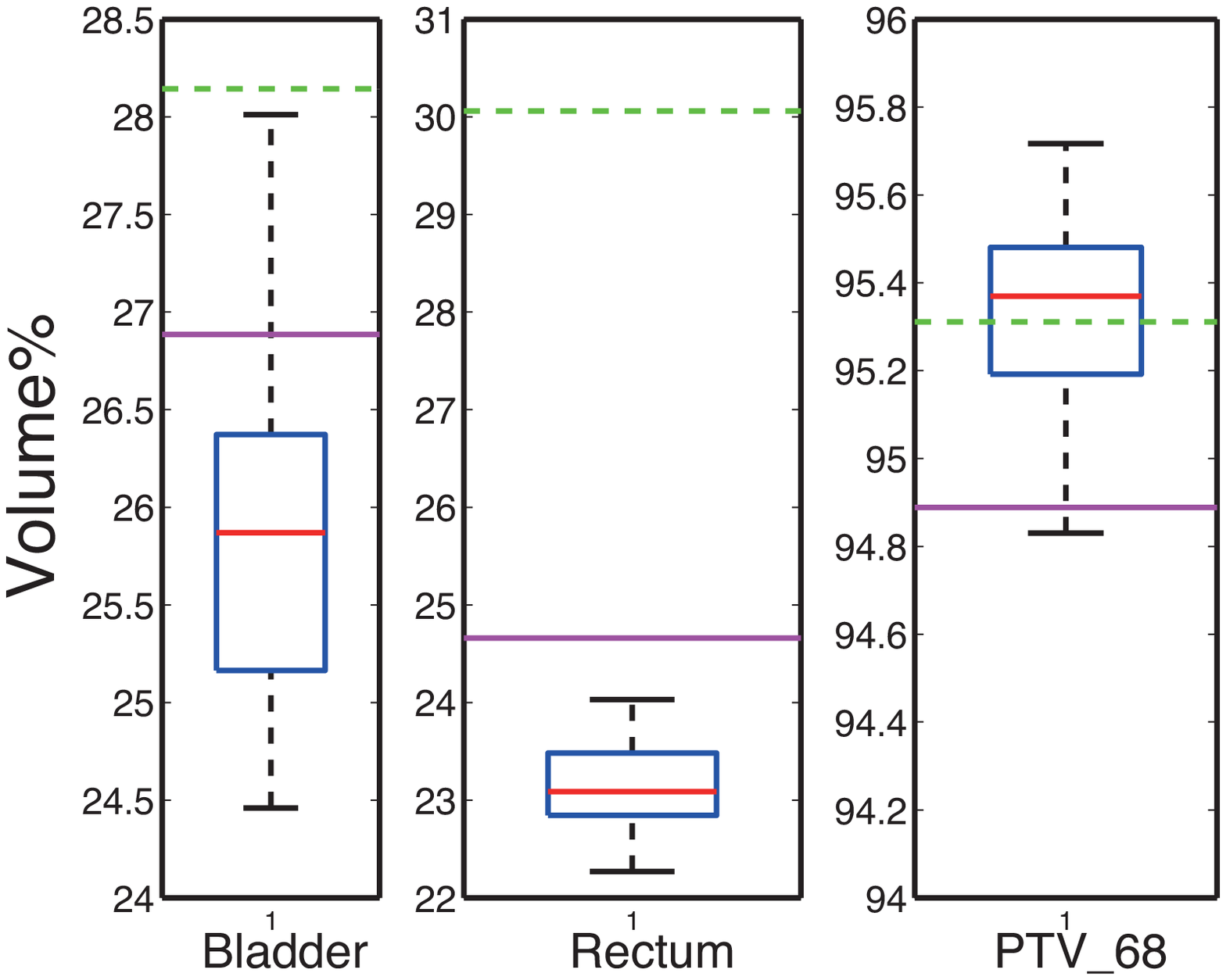}
\end{minipage}
\caption{
%Left: Computational results of 20 runs by RGAltMin on the  prostate case.  Right: DVH of 20 runs by RGAltMin on the prostate case.  The green dashed lines and the purple solid lines represent the corresponding values yielded by GEAltMin and GGAltMin respectively.  Since the runtime of the GEAltMin is much longer, it's omitted in the figure.
Boxplots of the 20 runs of RGAltMin on the prostate case. The
figures from left to right are corresponding to energy, runtime and DVHs of
Bladder, Rectum and PTV\_68, respectively.
  The green dashed lines and the purple solid lines
 represent the corresponding values yielded by GEAltMin and GGAltMin,
respectively. The runtime of the GEAltMin is not shown in the second
figure since it is way larger than 600 seconds.
}\label{fig:9}
\end{figure}

\begin{figure}[htp]
\centering
\subfigure{
\begin{minipage}[t]{0.32\linewidth}
\centering
\includegraphics[width=\textwidth]{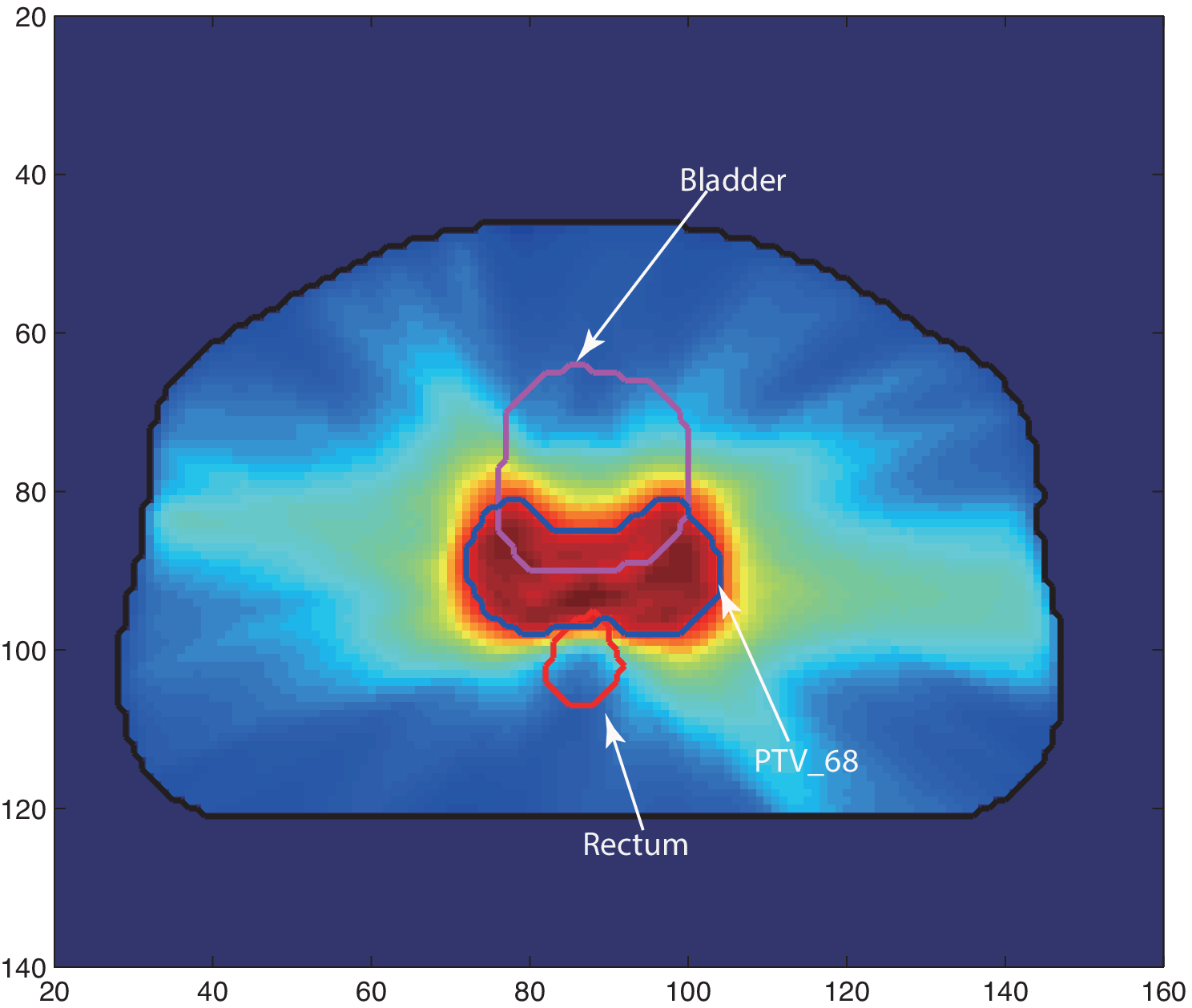}
\end{minipage}
\begin{minipage}[t]{0.32\linewidth}
\centering
\includegraphics[width=\textwidth]{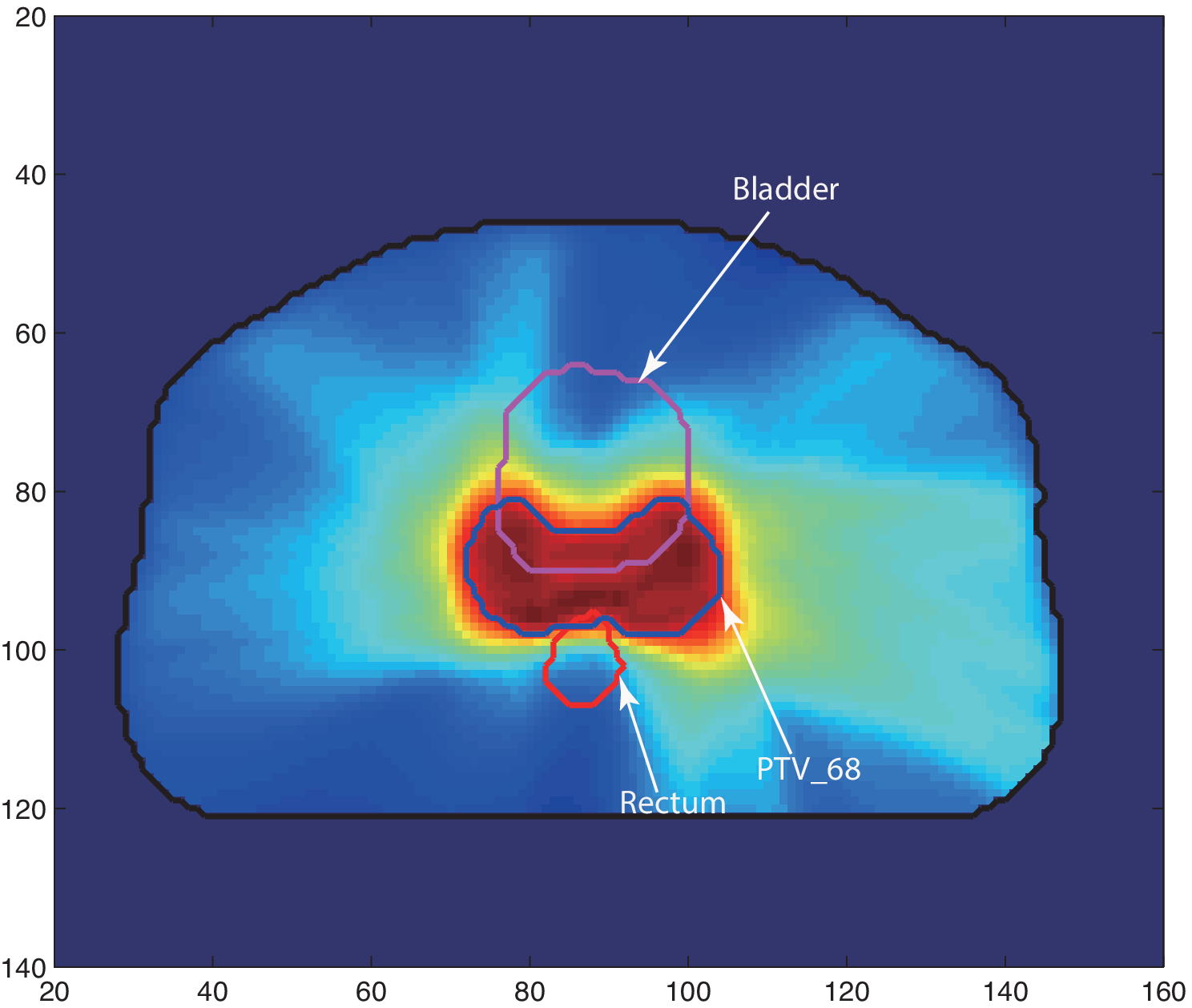}
\end{minipage}
\begin{minipage}[t]{0.32\linewidth}
\centering
\includegraphics[width=\textwidth]{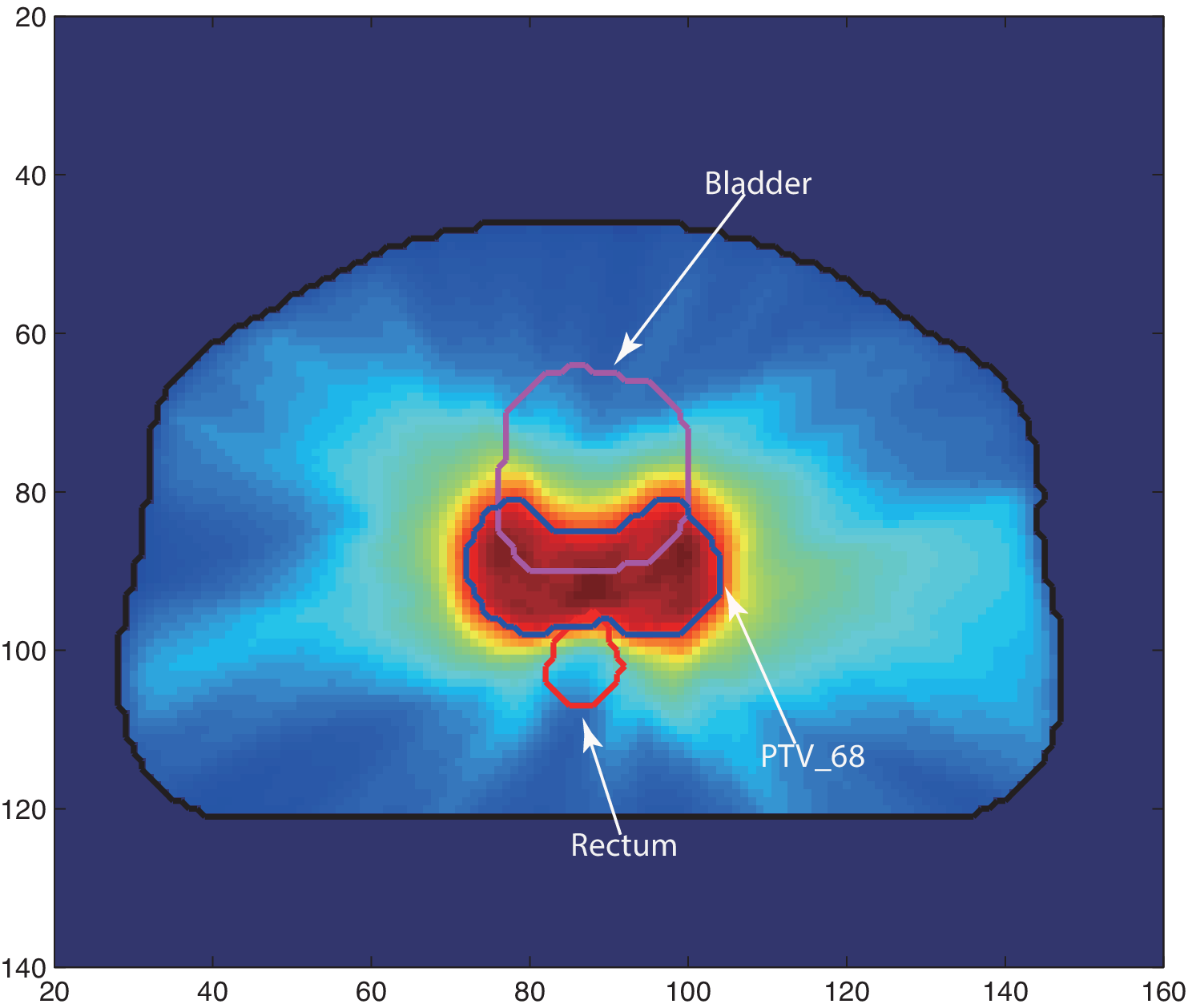}
\end{minipage}}
\subfigure{
\includegraphics[width=1\textwidth]{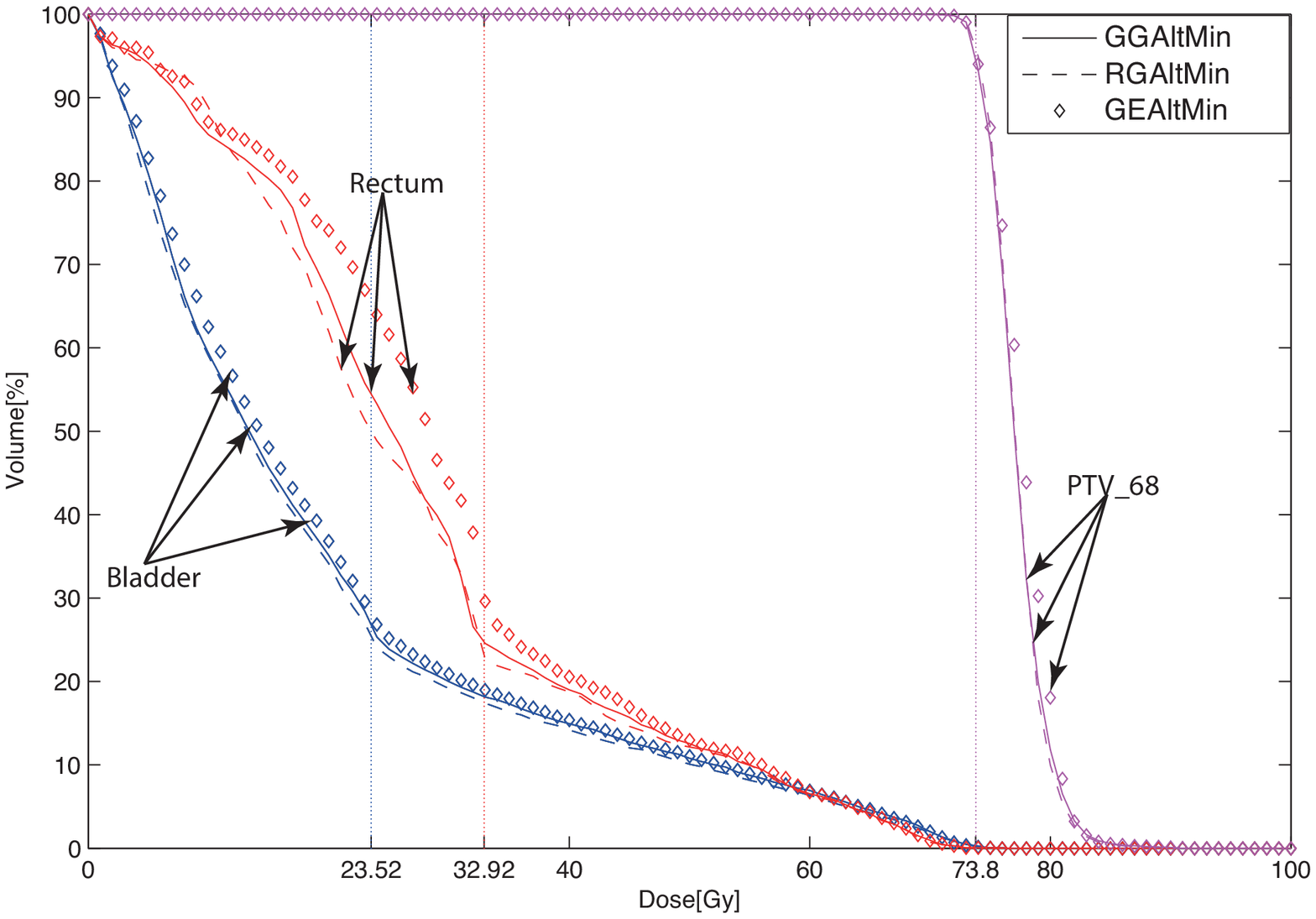} }
\caption{ Computational results on the prostate case.
The three figures from left to right on the first row correspond to the dose
distribution and their level-sets overlaying the outlined prostate and other
organs on a chosen slice of a patient's scan
 obtained by GGAltMin,
 RGAltMin and GEAltMin, respectively.  The second row shows the DVH curves for
GGAltMin (solid lines),
 RGAltMin (dashed lines) and GEAltMin (lines with diamond markers).
}\label{fig:6}
\end{figure}

Figure \ref{fig:6} shows the comparisons between the three algorithms. The dose distribution and their level-sets overlaying the outlined prostate and other
organs  on a particular slice of the patient's scan is  given in the first row of this
 figure. The red color means that the dose amount
 is high and the blue color suggests that the dose amount is low.
 The second row of this figure depicts the DVH curves. A summary of DVH is
 further shown in Table
 \ref{table2}.
 In general, the DVH should be low on the critical strucutures
Bladder and rectrum. The smaller these values are, the better the algorithm is.
On the other hand, a higher DVH is desired on the target PTV\_68.  We can observe that the dosage is high on target
 structures and it is low on critical structures. In particular, the curve of RGAltMin shows
 that 95.35\% of the target is dosed at or above the required level, and
74.19\% of the bladder and 76.70\% of the rectum are dosed below their safe limit in critical structures. The curve of GGAltMin shows that  94.87\% of the target is dosed at or above the
required level and   73.17\%
of the bladder and 75.36\% of the rectum are dosed below their safe limit in critical structures.
For GEAltMin, only 70\%  of the bladder and 75\% of the rectum are dosed below the safe limit respectively.
 The computational results of
the 20 runs of GEAltMin are further shown in Figure \ref{fig:9} using box plots. GEAltMin are better
than the other two algorithms in most cases and it is highly competitive to GGAltMin even
in the worst case.
%Boxplot of the 20 runs in Figure \ref{fig:9} shows that GEAltMin can outperform the other two algorithms in most cases and even the worst results are highly compatitive to GGAltMin.
 These figures and tables verify that  the randomized algorithm RGAltMin performs
 best in terms of the DVH quality and it is robust.
Finally, some samples of the aperture shapes and beam intensities for the VMAT plans
computed by GGAltMin is shown in the left side of Figure \ref{fig:8}.
% \revise{(Why not show the results of RGAltMin since it is the best algorithm?  Same for the other example).}

%From the  image of the dosage distribution, we can make sure that high dose is delivered to the target, while the critical structures receive low doses.

\begin{table}[htp]
\centering
\begin{tabular}{cccc}
\hline
algorithm &iteration&E& runtime\\ \hline
GEAltMin &$ 200$&\quad   $8.50e8$ & \quad $51$m$1$s\\
GGAltMin &$ 43$&\quad   $8.24e8$ &\quad  $9$m$33$s\\
RGAltMin &$ 64$&\quad   $8.20e8$ &\quad  $8$m$17$s\\
\hline
\end{tabular}
\caption{A summary of computational results on the prostate case. }\label{table1}
\end{table}

\begin{table}[htp]
\centering
\begin{tabular}{cccc}
\hline
 algorithm &\quad  Bladder&\quad  Rectum&\quad   PTV\_68\\ \hline
GEAltMin&\quad  $28.20\%$&\quad   $30.34\%$ &\quad  $95.33\%$\\
GGAltMin&\quad  $ 26.83\%$&\quad   $24.64\%$ &\quad  $94.87\%$\\
RGAltMin&\quad  $ 25.91\%$&\quad   $23.30\%$ &\quad  $95.35\%$\\
\hline
\end{tabular}
\caption{A summary of DVH on the prostate case. }\label{table2}
\end{table}

The second test is on the head-and-neck case, which is much more complicated
than the prostate case.  It involves two different targets, labeled as PTV\_70 and
PTV\_63, respectively. PTV\_70 is expanded on the GTV.  It consists of two
components of the tumors, and a dose of 70 Gy is prescribed. PTV\_63 is a larger target area which contains nearby high-risk region. A dose of 63 Gy is prescribed for this target. Two critical structures are the parotid and spinal cord PRV,
which have dose limits of 40Gy and 45Gy, respectively.  The head-and-neck case is complicated because of the number of targets, the
relative positions as well as the volumes of the target and critical structures. The parotid is close to PTV\_63 and much smaller compared to PTV\_63. If the DVH of PTV\_63 is increased, the dose on parotid will also be increased inevitably. Due to the small size of the parotid, violating its dose limit may only slightly increase the value of the energy function. Hence, an algorithm may sacrifice the parotid and fail to yield the desired result. If the penalty parameter
$\beta_r$ for the parotid is set large enough, it will also decrease the dosage received by PTV\_63. Therefore, this case is rather challenging.

There are  $1983$ angles in the original data of this case and its size is more than $60$GB. Since it may not be necessary to consider all the angles to obtain
good results, we only select 180 of the angles to cover all the angles when the couch angle is zero. A summary of computational results is reported in Table \ref{table3}. The results of RGAltMin are the average of 20 runs.
Table \ref{table3} shows that our algorithms converge after fewer iterations with much less time, and reach a smaller value of the energy function.  In particular, the randomized algorithm RGAltMin is significantly faster than GGAltMin. This demonstrates that introducing randomness into the alternating minimization framework is indeed helpful.

\begin{figure}[H]
\centering
\begin{minipage}[t]{0.48\linewidth}
\centering
\includegraphics[width=\textwidth]{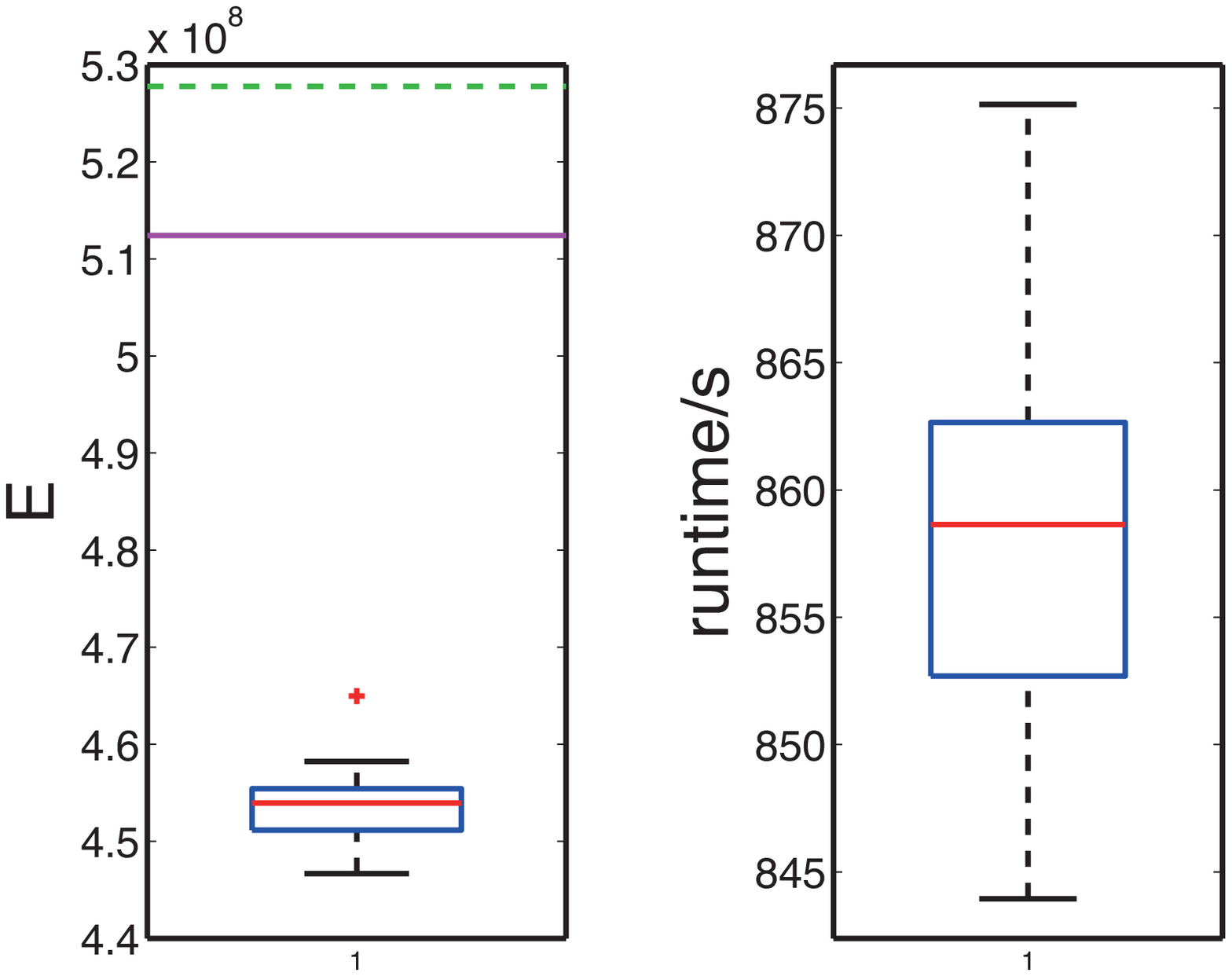}
\end{minipage}
\begin{minipage}[t]{0.48\linewidth}
\centering
\includegraphics[width=\textwidth]{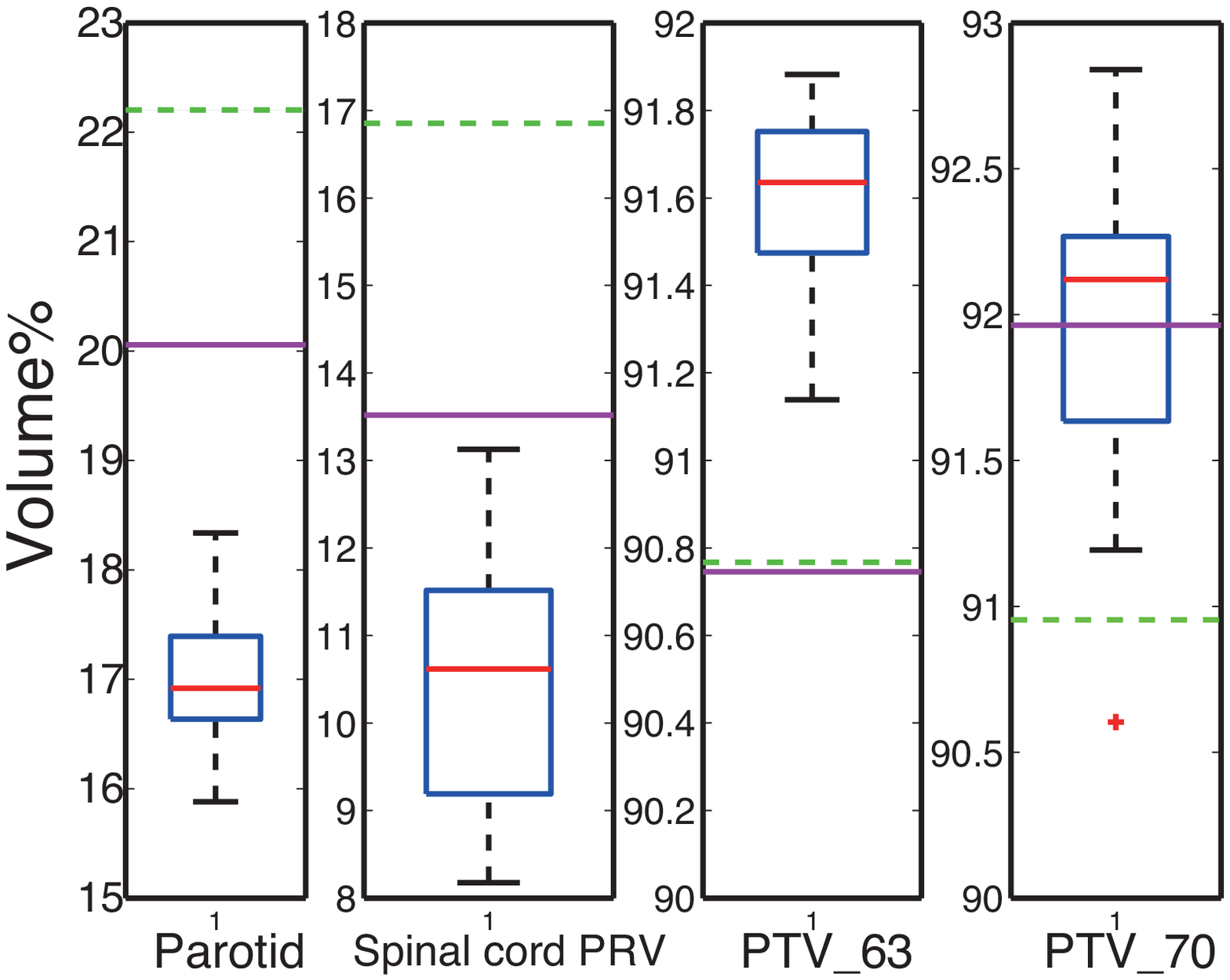}
\end{minipage}
\caption{
%Left: Computational results of 20 runs by RGAltMin on the  head-and-neck case.  Right: DVH of 20 runs by RGAltMin on the head-and-neck case.  The green dashed lines and the purple solid lines represent the corresponding values yielded by GEAltMin and GGAltMin respectively.  Since the runtimes of the other two algorithms are much longer, they are omitted in the figure.
Boxplots of the 20 runs of RGAltMin on the  head-and-neck case. The figures from left to right are corresponding to energy, runtime and DVHs of Parotid, Spinal cord PRV, PTV\_63 and PTV\_70, respectively.
The green dashed lines and the purple solid lines represent the corresponding
values yielded by GEAltMin and GGAltMin, respectively.  The runtimes of the
GEAltMin and GGAltMin are not shown in the second figure since they are way larger
than 875 seconds.
}\label{fig:10}
\end{figure}

\begin{figure}[htp]
\centering
\subfigure{
\begin{minipage}[t]{0.32\linewidth}
\centering
\includegraphics[width=\textwidth]{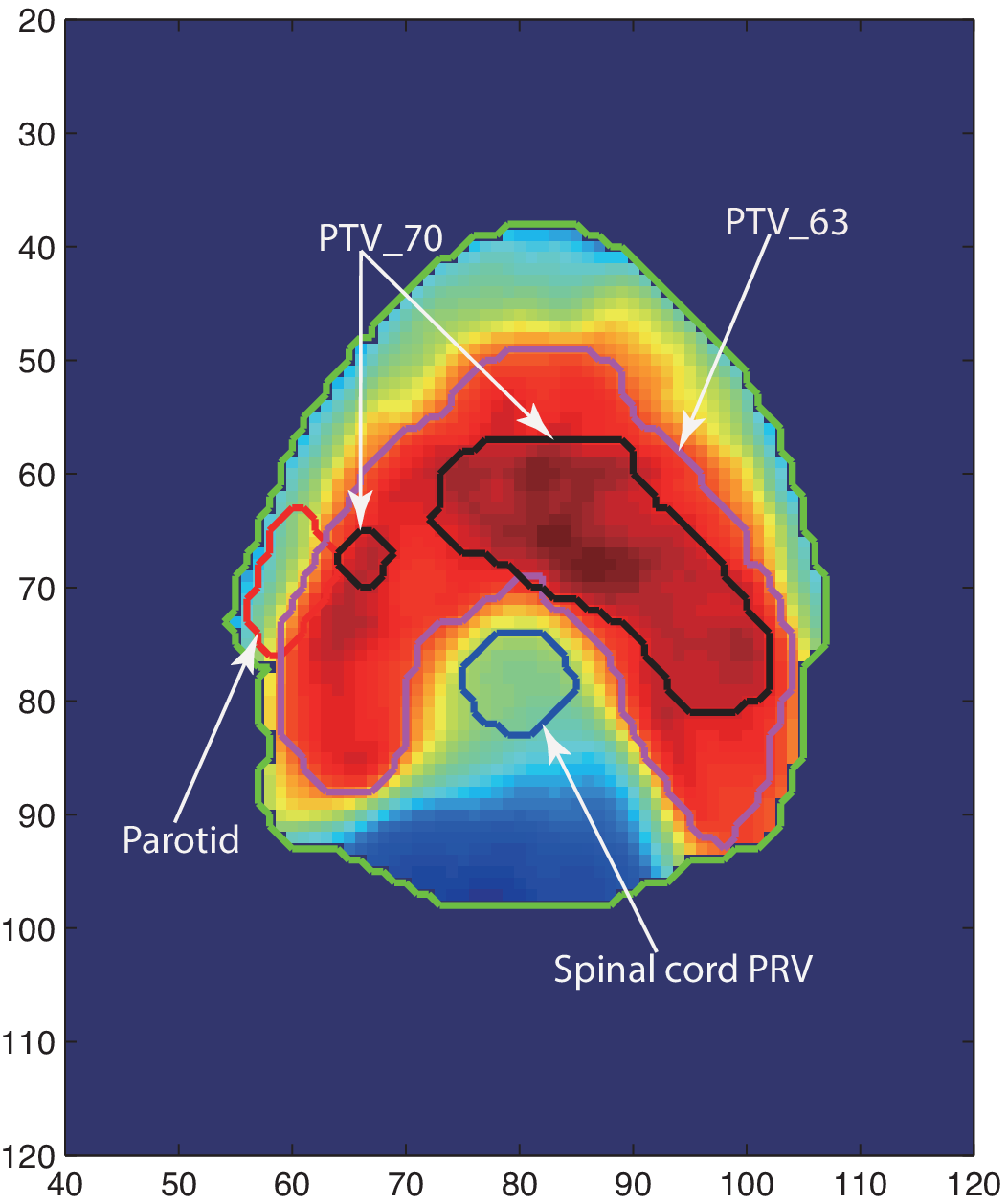}
\end{minipage}
\begin{minipage}[t]{0.32\linewidth}
\centering
\includegraphics[width=\textwidth]{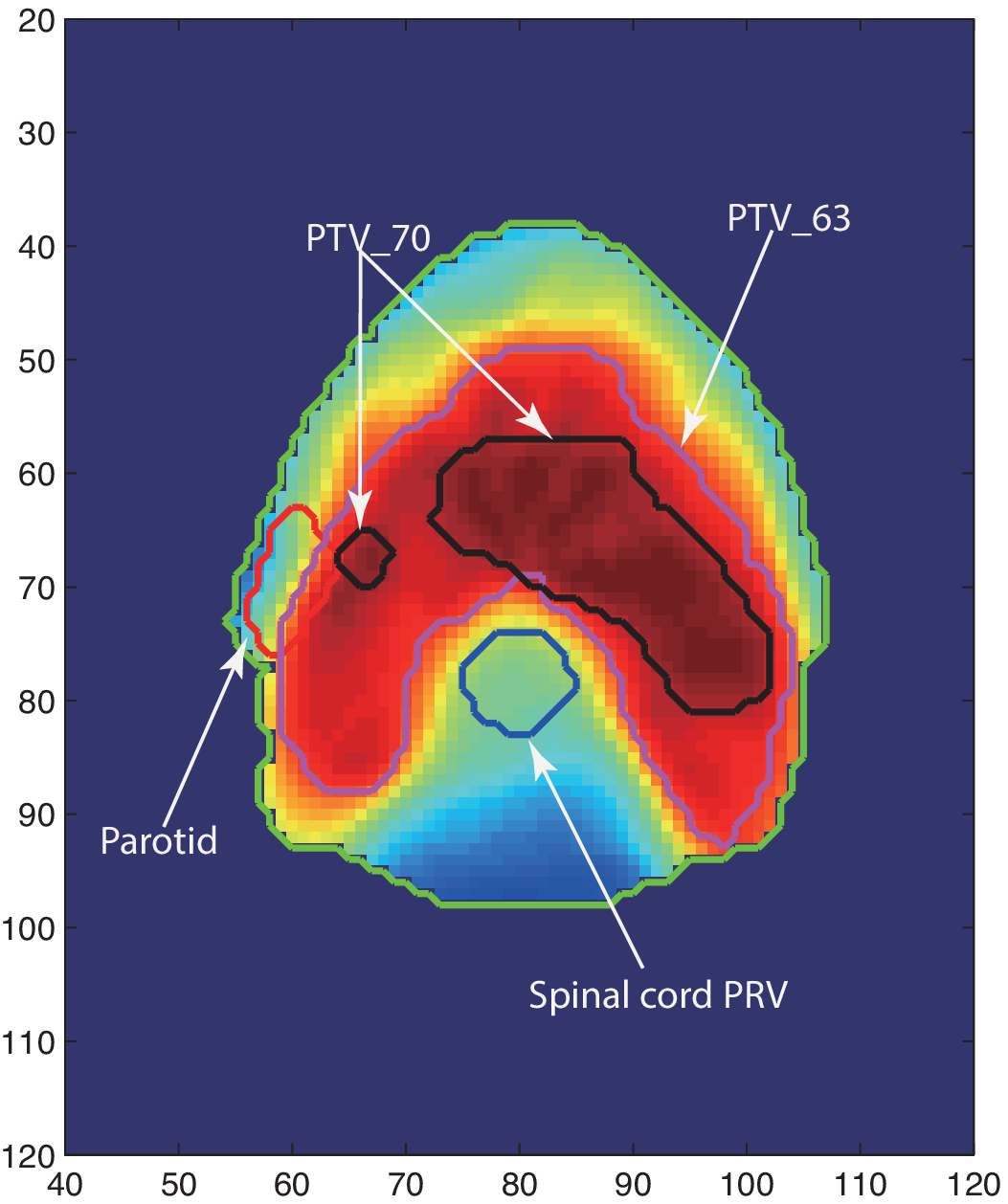}
\end{minipage}
\begin{minipage}[t]{0.32\linewidth}
\centering
\includegraphics[width=\textwidth]{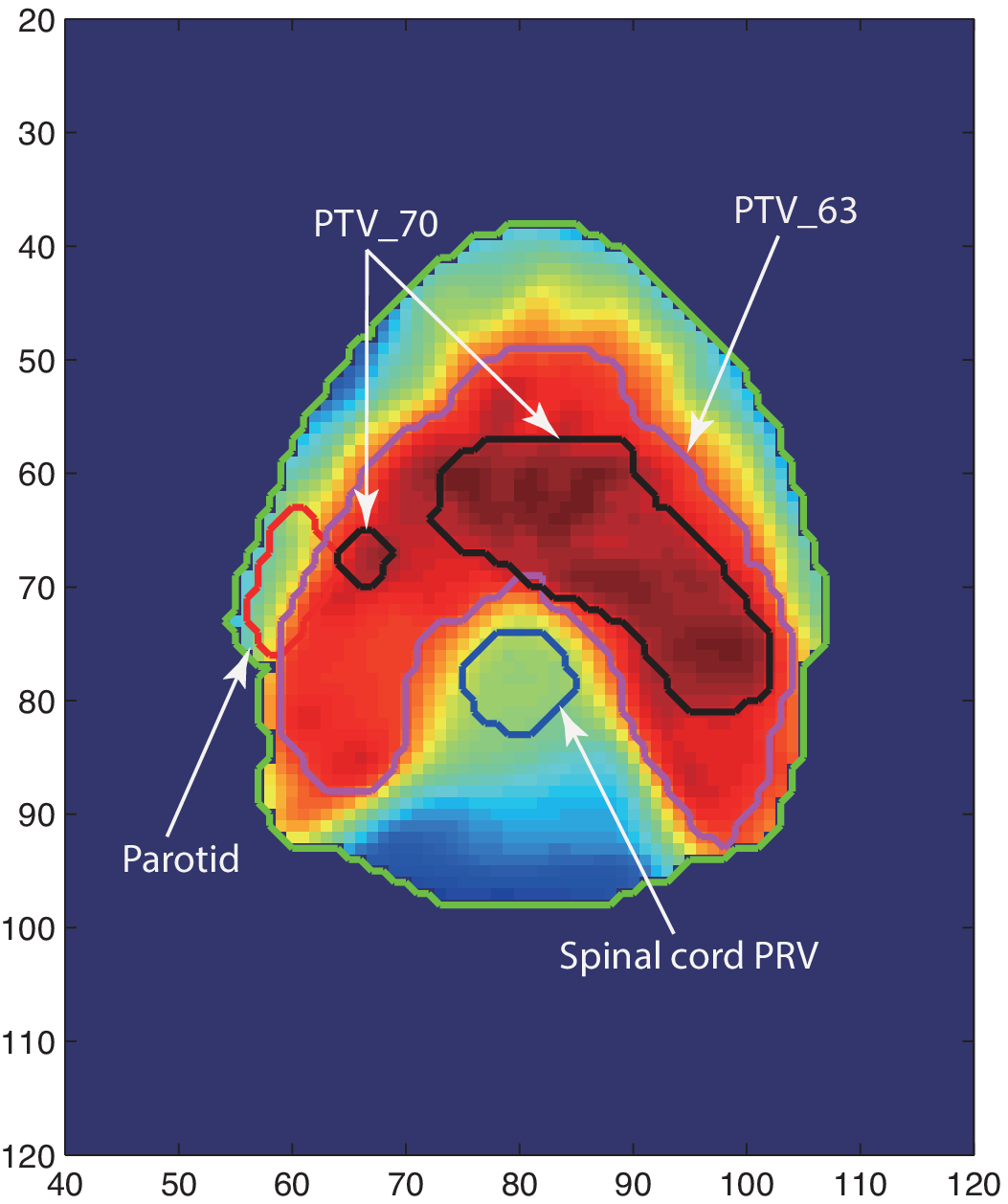}
\end{minipage}}
\subfigure{
\includegraphics[width=1.0\textwidth]{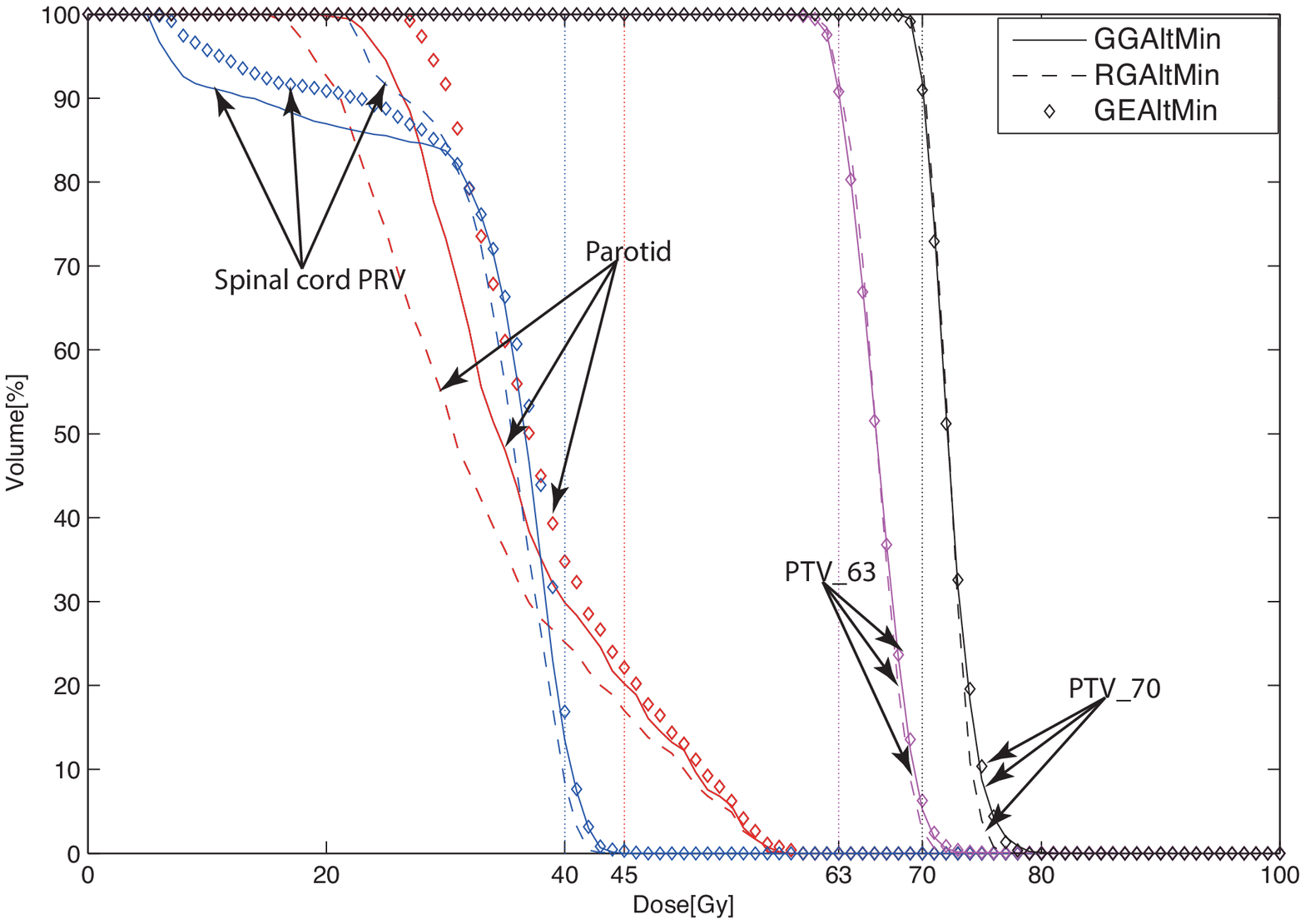} }
\caption{
 Computational results on the  head-and-neck  case.
The three figures from left to right on the first row correspond to the dose
distribution and their level-sets overlaying the outlined prostate and other
organs on a chosen slice of a patient's scan
 obtained by GGAltMin,
 RGAltMin and GEAltMin, respectively.  The second row shows the DVH curves for
GGAltMin (solid lines),
 RGAltMin (dashed lines) and GEAltMin (lines with diamond markers).
}\label{fig:7}
\end{figure}

\begin{figure}[htp]
%\setcaptionwidth{0.9\linewidth}
\centering
\subfigure{
\begin{minipage}[t]{0.5\linewidth}
\centering
\includegraphics[width=\textwidth]{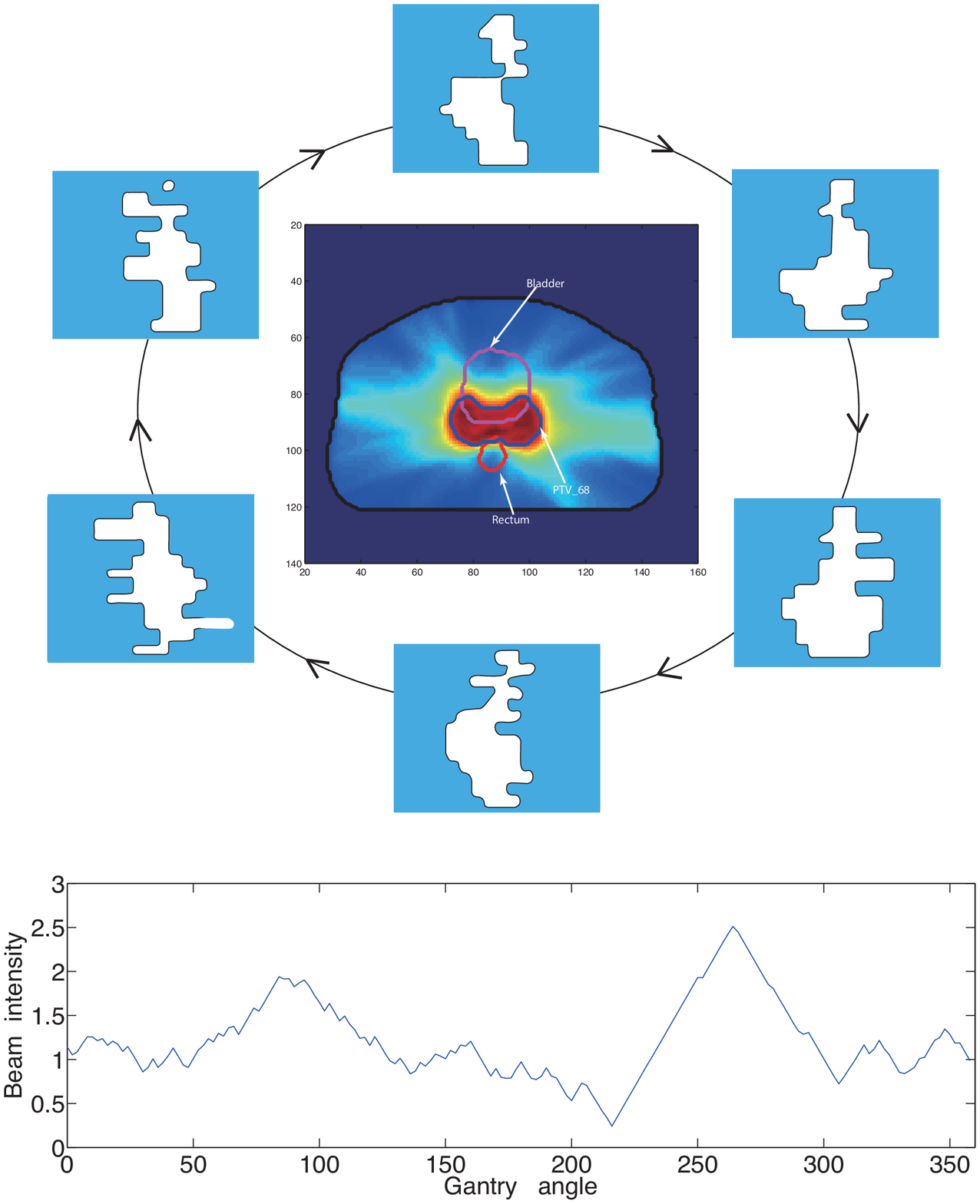}
\end{minipage}
\begin{minipage}[t]{0.5\linewidth}
\centering
\includegraphics[width=\textwidth]{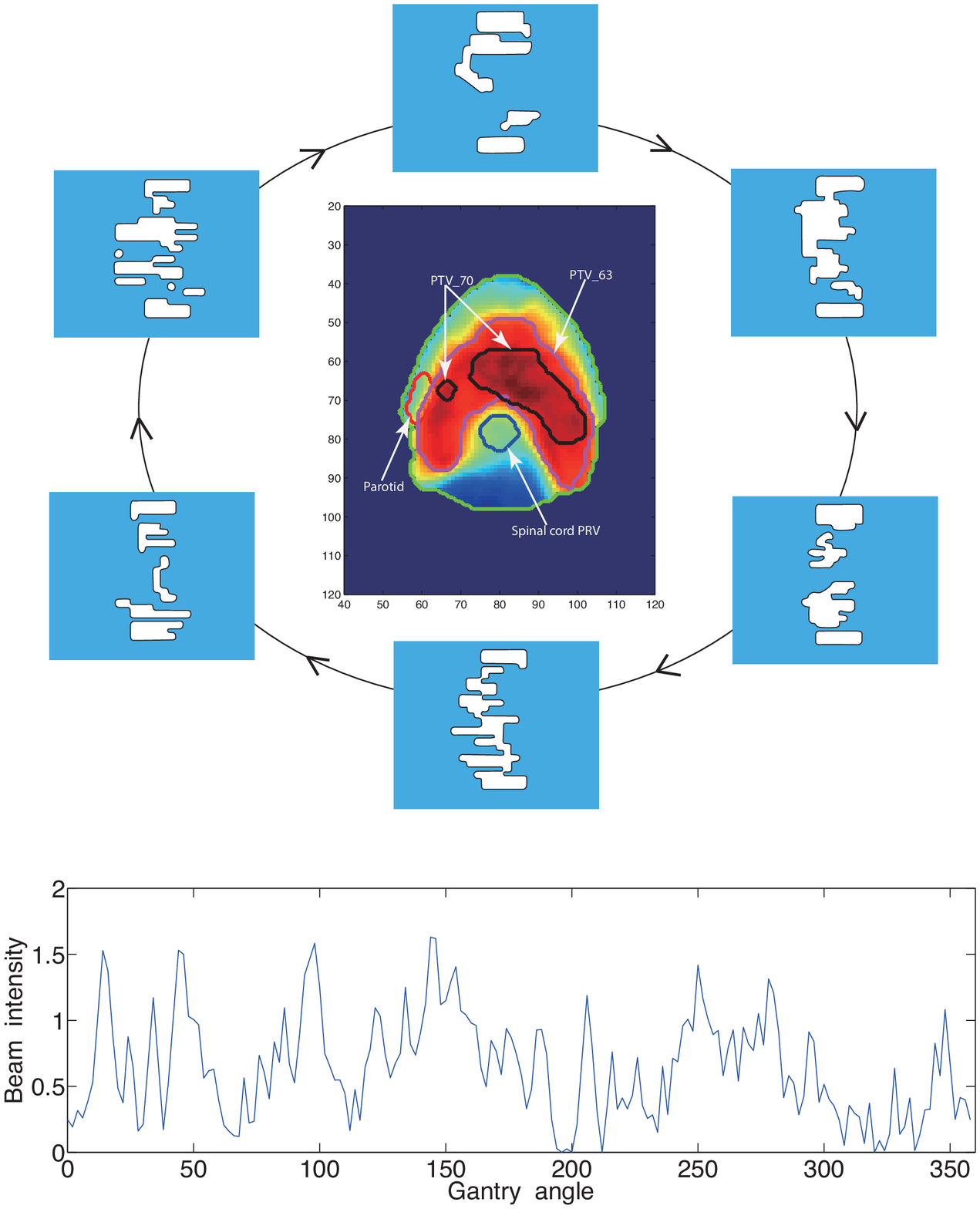}
\end{minipage}}
\caption{
A sample of the aperture shapes and beam intensities for the VMAT plans
computed by GGAltMin in the case of prostate cancer (left) and head-and-neck
cancer (right), respectively.
}\label{fig:8}
\end{figure}

The dose distribution and their level-sets overlaying the outlined prostate and other organs on a chosen slice of a patient's scan as well as the DVH curves are plotted in the first of Figure \ref{fig:7}. A summary of DVH is presented in Table \ref{table4}.  From these images and tables, the desired dosage distribution pattern are achieved by all algorithms.  In particular, the GGAltMin curve shows that  90.75\% of the PTV\_63 and 91.97\% of the PTV\_70 are dosed at or above the required level, and  79.77\% of the parotid and  86.50\% of the spinal cord PRV are dosed below their safe limit in critical structures. The RGAltMin curve shows that  91.59\% of the PTV\_63 and 91.97\% of the PTV\_70 dosed at or above the required level, 83.02\% of the parotid and  89.38\% of the spinal cord PRV dosed below their safe limit. Hence, our algorithms can yield the desired characteristics of a good treatment plan with sufficient doses applied to target structures and safe doses applied to critical structures.
%As shown in Figure \ref{fig:10}, even in the worst cases,  GEAltMin has much shorter runtime, smaller energy $E$ and less dose to critical structures. So the worst results yielded by GEAltMin are better than those by the other two algorithms.
The computational results of the 20 runs of GEAltMin are further shown in Figure \ref{fig:10}  using box plots. Even in the
worst case, GEAltMin consumes a much shorter runtime and reaches a smaller
energy E and less dose on the critical structures than the other two algorithms.
Hence, the performance of GEAltMin is quite robust.  Finally, the right side of Figure \ref{fig:8} shows some samples of the aperture shapes and beam intensities calculated by GGAltMin.

%The results of the two cases show that  Compared with the baseline algorithm, critical structures receive less dosage in our algorithms. What's more, our algorithms, especially algorithm 5, show a noticeable superiority in the runtime,  which proves the random sampling methods work quite well.

%The result of algorithm 5 is the mean value of 20 experiments. Parotid and spinal cord PRV are two critical structures, it's better for them to have a low DVH level.  However, for the targets PTV\_63 and  PTV\_70, a higher DVH is desired.  Thus from the number in the table, we can tell that our algorithms have better DVH quality.

\begin{table}[htp]
\centering
\begin{tabular}{ccccc}
\hline
algorithm &iteration&E&runtime&\\ \hline
GEAltMin & $300$& $5.28e8$ &$83$m$20$s&\\
GGAltMin &$163$& $5.13e8$ & $38$m$11$s&\\
RGAltMin & $ 71$&$4.55e8$ &$14$m$24$s&\\
\hline
\end{tabular}
\caption{A summary of computational results on the  head-and-neck case.
}\label{table3}
\end{table}

\begin{table}[htp]
\centering
\begin{tabular}{ccccc}
\hline
algorithm &\quad  ~Parotid&\quad  ~~Spinal cord PRV& \quad  PTV\_63&\quad
PTV\_70\\ \hline
GEAltMin&\quad  ~$22.12\%$&$16.88\%$ &\quad  ~$90.77\%$&\quad  ~$90.97\%$\\
GGAltMin&\quad  ~$ 20.23\%$&$13.50\%$ & \quad  ~$90.75\%$&\quad  ~$91.97\%$\\
RGAltMin&\quad  ~$ 16.98\%$&$10.63\%$ & \quad  ~$91.59\%$&\quad  ~$91.97\%$\\

\hline
\end{tabular}
\caption{A summary of DVH on the head-and-neck case.
}\label{table4}
\end{table}

\section{Conclusion} \label{sec:conclusion}
The optimization of VMAT treatment plans in cancer radiotherapy is an important problem in clinical application. The treatment plan optimization is complicated due to practical constraints imposed by the equipments involved. It requires to determine suitable aperture shapes and beam intensities in order to generate
sufficient doses applied to target structures while minimizing doses applied to critical structures as much as possible. In this paper, we consider constraints on the bounds on the beam intensity and its rate of change, limit on the moving speed of the leaves of the multi-leaf collimator (MLC), and the direction-convexity of the MLC. We propose a mixed-integer nonlinear and nonconvex model in discrete setting where the aperture shapes are characterized by integer variables to conveniently describe the directional convexity properties. The model is solved by performing an alternating minimization with respect to the aperture shapes and the beam intensity, separately. The aperture shapes are computed using a greedy strategy which is further enhanced by random sampling. The beam intensities are computed by a standard gradient-projection method using nonmonotone line search. Since calculating the dose distribution and evaluating the energy function are computationally expensive due to the large number of voxels in the discrete setting, we propose an incremental randomized strategy which only computes a proportion of the voxels at a time.  Comparing to the classical incremental algorithms, our selection of the voxels are based on probabilities defined by the importance of the voxels. Numerical simulations on the prostate and head-and-neck data sets confirm that our method is highly competitive to the state-of-the-art algorithms in terms of both computation efficiency and quality of treatment planning. 

The performance of our algorithms can be further improved in several ways, such as speeding up convergence and improving accuracy, with the help of the
recent techniques on mixed-integer programming and randomized optimization. Three particularly important topics for future investigations are (i) a comprehensive study of mathematical modelling of treatment planning in VMAT or other radiotherapy techniques,  (ii) parallelizing the computation in the greedy approach and random sampling, and (iii) extensive numerical experiments on clinical data sets.

\section*{Acknowledgements} We would like to thank Dr. Xun Jia from Department of Radiation Oncology, University of Southwestern Medical Center for his valuable inputs on this project and the clinical data sets. We would also like to extend our gratitude to Dr. Li-Tien Cheng for sharing his source codes of the algorithms in \cite{Dong2012}.

\bibliographystyle{siam}
\bibliography{VMAT}

\end{document}